\pdfoutput=1
\documentclass[preprint]{elsarticle}
\makeatletter
\def\ps@pprintTitle{%
 \let\@oddhead\@empty
 \let\@evenhead\@empty
 \def\@oddfoot{\centerline{\thepage}}%
 \let\@evenfoot\@oddfoot}
\makeatother
\usepackage{graphicx}
\usepackage{ amssymb }
\usepackage{ amsmath }
\usepackage{ amsfonts}
\usepackage{ enumerate}
\usepackage{wrapfig}
\usepackage{ appendix }
\biboptions{sort&compress}
\usepackage{ float }
\usepackage{caption}
\usepackage{subcaption}
\let\OLDthebibliography\thebibliography
\renewcommand\thebibliography[1]{
  \OLDthebibliography{#1}
  \setlength{\parskip}{0pt}
  \setlength{\itemsep}{0pt plus 0.3ex}
}

\usepackage{color}
\usepackage{todonotes}
\usepackage{soul}
\newcommand{\R}{\mathbb{R}}
\newcommand{\C}{\mathbb{C}}
\newcommand{\E}{\mathbb{E}}

\newcommand{\eps}{\epsilon}
\newcommand{\bra}{\langle}
\newcommand{\ket}{\rangle}
\newcommand{\bm}{\boldsymbol}

\usepackage{geometry}
\begin{document}

\begin{frontmatter}
\title{Solving the stochastic Landau-Lifshitz-Gilbert-Slonczewski equation for monodomain nanomagnets : A survey and analysis of numerical techniques}

\author[add2]{Sebastian Ament\corref{cor1}}
\ead{sebastian.ament@nyu.edu}

\author[add1]{Nikhil Rangarajan}
\ead{nikhil.rangarajan@nyu.edu}

\author[add1]{Arun Parthasarathy}
\ead{arun.parth@nyu.edu}

\author[add1]{Shaloo Rakheja}
\ead{shaloo.rakheja@nyu.edu}

\cortext[cor1]{Corresponding author}

\address[add1]{Department of Electrical Engineering, New York University, NY 11201}
\address[add2]{Courant Institute of Mathematical Sciences, New York University, NY 10012}

\begin{abstract}

The stochastic Landau-Lifshitz-Gilbert-Slonczewski (s-LLGS) equation is widely used by researchers to study the temporal evolution of the macrospin subject to spin torque and thermal noise. The numerical simulation of the s-LLGS equation requires an appropriate choice of stochastic calculus and numerical integration scheme.  In this paper, we first comprehensively evaluate the accuracy and complexity of various numerical techniques to solve the s-LLGS equation. We focus on implicit midpoint, Heun, and Euler-Heun methods that converge to the Stratonovich solution of the s-LLGS equation. By performing several numerical tests, for both strong (path-wise) and weak (statistical) convergence, we quantify the accuracy of various numerical schemes used to solve the s-LLGS equation. We demonstrate a new method intended to solve stochastic differential equations (SDEs) with small noise (the RK4-Heun method), and test its capability to handle the s-LLGS equation. We also discuss the circuit implementation of nanomagnets for large-scale SPICE-based simulations in a standard hierarchical circuit simulator. We evaluate the efficacy of SPICE in handling the stochastic dynamics of the multiplicative noise in the s-LLGS equation. Numerical schemes such as Euler and Gear, which are typically used by SPICE-based circuit simulators do not yield the expected outcome when solving the Stratonovich s-LLGS equation. While the trapezoidal method in SPICE does solve for the Stratonovich solution, its accuracy is limited by the minimum time step of integration in SPICE. We implement the s-LLGS equation in both its cartesian and spherical coordinates form in SPICE and compare the stability and accuracy of the two implementations. The results in this paper will serve as guidelines for researchers to understand the tradeoffs between accuracy and complexity of various numerical methods and the choice of appropriate calculus to solve the s-LLGS equation.
\end{abstract}


\end{frontmatter}

\section{Introduction}
\label{sec:introduction} 
Beyond-CMOS technologies using alternate state variables, such as electron spin, magnetic domains, spin waves, phase change, and photons have garnered significant interest lately for low-power and robust computing applications \cite{galatsis2009alternate, allwood2005magnetic, slonczewski1999excitation, qureshi2009scalable}. Spin-based devices, implemented with nanomagnets, have become an extrememly popular choice for magnetic memory, storage, and logic owing to their non-volatility, low power, and low area footprints. Spintronics has established itself as a front-runner in this race of beyond-CMOS technologies, with devices such as the spin-transfer torque (STT) MRAM (that offers high density and high-performance non-volatile storage coupled with low power and cost) predicted to replace conventions DRAMs in the near future  \cite{lin200945nm, huai2008spin, kultursay2013evaluating}. The dynamics and performance of the majority of spin-based devices are modeled using the stochastic Landau-Lifshitz-Gilbert-Slonczewski (s-LLGS) equation. This equation describes the temporal evolution of the magnetization vector, $\bm{M}$, of a monodomain nanomagnet under the effects of magnetic fields, STT and thermal noise  \cite{Slonczewski1996, Ralph, Stiles, Liz, stiles2006}. Mathematically, the s-LLGS equation is given as
\begin{equation}
\frac{d\bm{M}}{dt} = -\gamma\mu_0(\bm{M}\times \bm{H}_{eff}) + \frac{\alpha}{M_s} \Big(\bm{M}\times\frac{d\bm{M}}{dt}\Big) - \frac{\bm{M} \times (\bm{M} \times \bm{I_s})}{qN_sM_s},  
\label{basic_sLLGS}
\end{equation}
\noindent where $\gamma$ is the gyromagnetic ratio of electron, $\bm{H}_{eff}$ is the effective magnetic field experienced by the nanomagnet, $\alpha$ is the dimensionless Gilbert damping constant, $M_s$ is the saturation magnetization of the nanomagnet, $\bm{I}_s$ is the applied spin current, $q$ is the elementary charge, $N_s$ is the number of spins given as $N_s = \frac{2M_s V}{\gamma\hbar}$, and $V$ is the volume of the nanomagnet.\\
\noindent The nanomagnets used in spintronic devices and memories today are sub-$100$ nm \cite{chun2013scaling, cascales2013magnetization}; for such small dimensions the macrospin approximation \cite{Tannous2006} validates the use of the monodomain model, wherein the nanomagnet is assumed to possess a single coherent magnetization ($\bm{M}$) over the entire domain and any spatial variation in $\bm{M}$ is not considered. This approximation is only valid for nanomagnets of dimensions smaller than the critical Stoner radius, which is of the order of a few $100$s of nm for typical magnetic materials \cite{Tannous2006}. The first term on the right hand side (RHS) in \eqref{basic_sLLGS} is the conservative precessional torque that governs the precession of the magnetization vector around the effective field acting on the nanomagnet. This effective field comprises the magnetocrystalline anisotropy field, the shape anisotropy field, and the external applied field. A Langevin field $\bm{h}_r = h_x \bm{\hat{x}} + h_y \bm{\hat{y}} + h_z \bm{\hat{z}}$, representing Gaussian white noise,
is added into the effective field in the s-LLGS equation to model thermal noise. 
The second term on the RHS in (\ref{basic_sLLGS}) is the Gilbert damping torque, 
which is responsible for damping the precessions of the magnetization vector
and eventually relaxing it to one of its stable states 
\cite{Aquino2004}. The final term on the RHS in \eqref{basic_sLLGS} is the Slonczewski spin torque arising from the deposition of spin angular momentum by the itinerant electrons of the spin-polarized current. The field-like torque \cite{xiao2005}, (FLT) which results from the non-equilibrium spin accumulation in the nanomagnet, where the transverse component of spin current may persist with a characteristic length of few angstroms or nanometers, is negligible compared to the STT and is not considered in this study. For analysis in this paper, we transform \eqref{basic_sLLGS} into its dimensionless form expressed as\footnote{The spherical coordinates representation of the s-LLGS equation is discussed in section 5.}

\begin{equation}
\label{eq_llg_implicit1}
\frac{d\bm{m}}{dt} =  - \bm{m} \times \bm{h}_{eff} + \alpha \left( \bm{m} \times \frac{d\bm{m}}{dt} \right) - \bm{m} \times ( \bm{m} \times \bm{i}_s) ,\\
\end{equation}

\noindent where we have the normalized quantities $\bm{m} = \frac{\bm{M}}{M_s}$, $\bm{h}_{eff}  = \frac{\bm{H}_{eff}}{M_s}$, and $\bm{i_s} = \frac{\bm{I}_s}{I}$. Here, the scaling factor, $I$, for spin current is defined as $I = q\gamma \mu_0M_s N_s$. The time scale is normalized using the factor $(\gamma \mu_0 M_s)^{-1}$. The advantages of the normalized equation \eqref{eq_llg_implicit1} over \eqref{basic_sLLGS} are: (a) it is easier to deal with normalized quantities in terms of numerical complexity, and (b) normalized entities are mathematically well behaved under the application of a numerical scheme. The explicit form of \eqref{eq_llg_implicit1} obtained by decoupling $d\bm{m}/dt$ is given as (see Appendix A for the detailed derivations)
 \begin{equation}
\label{eq_llg_explicit1}
\frac{d\bm{m}}{dt} = - \frac{1}{1+\alpha^2} \left[ \bm{m} \times \bm{h}_{eff} + (\bm{m} \times (\bm{m} \times \bm{i}_s)) + \alpha \left( \bm{m} \times( \bm{m} \times \bm{h}_{eff})) - \alpha (\bm{m} \times \bm{i}_s \right)\right].
\end{equation}

\noindent Prior works \cite{llg1,llg2,llg3} have reviewed numerical techniques to solve the LLG equation, but a clear and comprehensive treatment of the interplay of thermal noise and deterministic dynamics is lacking. Reference \cite{Aquino2006} applies the implicit midpoint integration technique to the deterministic LLG equation (without thermal noise) and details the intricacies of the accuracy, stability, and time step used in the simulation. But it falls short because all the physical systems in existence experience finite temperature effects; hence, there is a need to include the thermal energy term in the effective field acting on the nanomagnet. The addition of the thermal noise transforms the deterministic LLG equation into a stochastic differential equation (SDE), which warrants an appropriate and careful choice of stochastic calculus and numerical integration scheme for correct convergence. In Ref. \cite{banas2013computational}, the authors analyze the implicit midpoint scheme for a single spin, a finite ensemble of spins, and an infinite ensemble of ferromagnetic spins, but they do not consider the Slonczewski spin torque, which is essential since STT-based switching of nanomagnets has become immensely popular in recent devices owing to its efficiency and reliability.  Reference \cite{banas2013computational} also does not offer any comparison of the implicit midpoint scheme with other numerical schemes or present any quantitative evidence as to why one method must be preferred over others with respect to solving the s-LLGS equation.  Reference \cite{Pinna2013} presents the impact of thermal noise on the magnetization reversal, but it does not present the details of the numerical scheme employed to treat the thermal noise. In this research, we leverage these prior works to carefully study the impact of the specific numerical scheme to solve the s-LLGS equation on the macrospin evolution. We propose a new method called the RK4-Huen, which is particularly useful when the stochastic term in the SDE is much smaller than the deterministic term, which is usually the case in magnetic bodies. We also comment on the importance of the time step used for numerical integration on the accuracy of various numerical schemes. The second part of this paper focuses on SPICE simulations of nanomagnet dynamics. SPICE-based models are widely used to facilitate the design and optimization of complex magneto-logic networks. However, it is important to represent the s-LLGS equation in SPICE in a manner that the existing numerical integration schemes in SPICE can successfully handle the multiplicative white noise in the system. We quantify the performance (accuracy and stability) of SPICE solvers that use numerical schemes, such as trapezoidal and Euler, to solve the s-LLGS equation in both cartesian and spherical coordinate systems.\\
\noindent The remainder of this paper is organized as follows. Section~\ref{sec:eff_field} describes the formulation of the effective magnetic field acting on the macrospin. The details underlying the midpoint method are presented in Section~\ref{sec:midpoint}. In Section~\ref{sec:methods}, the accuracy of various numerical methods, including a method for small noise, to solve the s-LLGS equation is presented. We conclude the paper by comparing the results of the implicit midpoint method and those obtained from SPICE solvers and the NIST-standard micromagnetic tool OOMMF.

\vspace{-2ex}

\section{Description of the effective magnetic field}
\label{sec:eff_field}

To arrive at the expression for the effective field in the s-LLGS equation, 
we consider the following energies, per unit volume, in the energy landscape of the macrospin
 \cite{Bruno, Pinna2015,spaldin2010,2005Aquino}:
\begin{enumerate}[(1)]
\item Zeeman energy due to an externally applied field, $E_{zeeman} = -\mu_0 (\bm{H}_{app} \cdot \bm{M}),$
\vspace{-1ex}
\item Uniaxial anisotropy energy, $E_{uniaxial} = -K_u\cos^2\theta,$
\vspace{-1ex}
\item Shape anisotropy energy, $E_{shape} = \frac{1}{2}\mu_0 (N_xM_x^2 + N_yM_y^2 + N_zM_z^2),$
\vspace{-1ex}
\item Thermal energy, $E_T.$
\end{enumerate}

\noindent In the above set of equations, $\bm{H}_{app}$ is the external magnetic field, $K_u = \frac{1}{2}\mu_0M_sH_k $ is the uniaxial energy density ($H_k$ is the anisotropy field), $\theta $ is the angle between the easy axis and the magnetization $(\cos\theta = \bm{\hat{n}} \cdot \bm{m} = \bm{\hat{n}} \cdot \frac{\bm{M}}{M_s})$, and $N_x, N_y$ and $N_z$ are the geometry-dependent demagnetization coefficients of the nanomagnet. The exchange energy, which originates from the overlap of electron orbitals, is a very short range force that acts on the order of the exchange length of the magnetic material. Since the exchange lengths of typical magnetic materials are on the order of a few nm, and the size of the nanomagnets used in applications today have dimensions of few 10's of nm, the exchange energy becomes insignificant compared to  long range interactions like the dipolar interaction, that results in the shape anisotropy term \cite{pinna2015spin}. Hence, for the purposes of this paper, we neglect the exchange energy term in the formulation of the effective field.   \\
\noindent The thermal field $\bm{H}_T(t)$ can be expressed in terms of the Wiener process as $\bm{H}_T(t) dt = \nu d\bm{W}(t) $ \cite{Aquino2006}, where $\bm{W}(t)$ is the Wiener process, and $\nu =\sqrt{\frac{2\alpha K_bT}{\mu_0 M_s^2 V }}$ \cite{Sun2006, Sasikanth2012}. Here{\color{red}{,}} $K_bT$ is the thermal energy. The statistical properties of this thermal field discussed by Brown and Kubo are given as \cite{Brown1963, kubo1970}\\
\indent (1) The mean thermal field: $\bra \bm{H}_{T,i}(t) \ket= 0,$\\
\indent (2) The correlation between the components of $\bm{H}_T(t)$ defined over a time interval $\tau$, \begin{equation}
\bra \bm{H}_{T,i}(t)\bm{H}_{T,j}(t+\tau) \ket = \frac{2K_bT\alpha}{\gamma \mu_0^2 M_s V}\delta_{ij}\delta(\tau),
\end{equation}
where $\delta_{ij}$ is the Kronecker delta function. To simulate the thermal effects numerically, we discretize the model in time
\begin{equation}
\label{eq_discretized_thermal_model}
\bm{H}_T(t) \Delta t = \nu \Delta \bm{W}(t),
\end{equation}
where $\Delta \bm{W}(t) = \bm{W}(t + \Delta t) - \bm{W}(t)$. The normalized standard deviation of the thermal field is given by
\begin{equation}
\sigma = \sqrt{\frac{2\alpha K_bT}{\mu_0 M_s^2 V}}  \sqrt{ \frac{\Delta t' }{\gamma \mu_0 M_s} },
\end{equation}
where $\Delta t$ is the time step of the numerical method used and $t' = (\gamma \mu_0 M_s) t$.

\noindent We then have
\begin{equation}
\bm{h}_T(t) \Delta t' = \sigma \xi_t,
\end{equation}
where the normalized thermal field $\bm{h}_T = \bm{H}_T / M_s$, and $\xi_t \sim \mathcal{N}(0,1)$ is a standard Gaussian vector.\\

\noindent The total energy of the macrospin (excluding the thermal energy) is 
\begin{equation}
\begin{aligned}
E_{total}  &= V[E_{zeeman} + E_{uniaxial} + E_{shape}] \\
&= V\Big[-\mu_0 (\bm{H}_{app}\cdot \bm{M}) - K_u\cos^2\theta + \frac{1}{2}\mu_0 (N_xM_x^2 + N_yM_y^2 + N_zM_z^2)\Big].
\end{aligned}
\end{equation}

\noindent The total normalized effective field is then given as\\
\begin{equation}
\label{heff_1}
\begin{aligned}
\bm{h}_{eff} &= \frac{\bm{H}_{eff}}{M_s} = \frac{-1}{\mu_0 M_s V}\nabla_{\bm{M}}E_{total}(\bm{M}) + \bm{h}_T\\
 &= \bm{h}_{app} + \frac{H_k}{M_s}(\bm{\hat{n}} \cdot \bm{m}) \bm{\hat{n}} - \sum_i N_i\bm{m}_i +\bm{h}_T.,  
 \end{aligned}
 \end{equation}
 
\noindent where we have normalized as $\bm{h}_{app} = \frac{\bm{H}_{app}}{M_s}, \bm{m}_i = \frac{\bm{M}_i}{M_s},$ and $\nabla_{\bm{M}}$ is the gradient with respect to the magnetization $\bm{M}$.\\
\vspace{-15pt}
\section{The Implicit Midpoint method}\label{sec:midpoint}
Using a numerical scheme based on the midpoint rule ensures that~(\ref{eq_llg_implicit1}) converges to the Stratonovich solution in the limit of $\Delta t \rightarrow$ 0. 
In this section, we define the implicit midpoint method, discretize the deterministic LLGS equation, and derive the Jacobian for use in the Gauss-Newton algorithm.
\subsection{The deterministic case}
For simplicity, we start the discussion with a one-dimensional (1D) deterministic ordinary differential equations (ODE) of the form 
\[
y'(t) = f(t, y(t)),
\]
with initial condition $y(0) = y_0$. The implicit midpoint update is given as
\begin{equation}
\label{eq_implicit1}
y_{n+1} = y_n + \Delta t \cdot f\left( t_n + \frac{\Delta t }{2} , \frac{y_n + y_{n+1}}{2}\right ),
\end{equation}
where $t_0 = 0$ and $t_{n+1} = t_n + \Delta t$.
In general, \eqref{eq_implicit1} is a nonlinear equation of $y_{n+1}$. To solve it for $y_{n+1}$ in the 1D case, one can apply Newton's method on the system 
\begin{equation}
S(y) = y - y_n - \Delta t \cdot f\left( t_n + \frac{\Delta t }{2} , \frac{y_n + y}{2}\right ).
\end{equation}

\noindent In order to solve a non-linear system $\bm S:\C^n \to \C^m$, we can use a generalized version of Newton's method, the Gauss-Newton algorithm.
Analogous to the 1D case, we can Taylor expand a differentiable system $\bm S$
\begin{equation}
\bm S(x) = \bm S( \bm x_0) + J[\bm S](\bm x_0) \cdot (\bm x - \bm x_0) + \eps,
\end{equation}
where $J[\bm S](\bm x_0)$ is the Jacobian of $\bm S$ at $\bm x_0$. 
Note that $\bm x, \bm x_0 \in \C^n$, $\bm S(\bm x) \in \C^m$, and $J[\bm S](\bm x_0) \in \C^{m \times n}$, and $\eps = o(\|\bm x- \bm x_0\|_2^2)$.
The $ij^{\text{th}}$ entry of the Jacobian of $\bm S$ is defined by 
\begin{equation}
J[\bm S](\bm x)_{ij} = \frac{ \partial S_i( \bm x) }{\partial x_j},
\end{equation}
where $S_i$ is the $i^{\text{th}}$ component of $\bm S$, and $x_j$ is the $j^{\text{th}}$ component of $\bm x$.
Setting $\bm S( \bm x) = 0$, dropping the higher-order terms, and setting $\bm x_{n+1} = \bm x$, $\bm x_n = \bm x_0$, we can write
\begin{equation}
\bm x_{n+1} = \bm x_n - J[\bm S](\bm x_n)^{-1} \cdot \bm S(\bm x_n),
\end{equation}
if $J[\bm S]$ is invertible.
As a result, we can use implicit methods on multi-dimensional ODE's, including the s-LLGS equation.

\subsection{Discretizing the deterministic Landau-Lifshitz-Gilbert-Slonczewski equation}
In order to apply the midpoint method, we need to discretize \eqref{eq_llg_implicit1}. Note that \eqref{eq_implicit1} takes the form
\begin{equation}
\label{eq_llg_implicit_step}
\bm m_{n+1} = \bm m_n + \Delta t \cdot \bm f(t_n + \Delta t / 2, \bm m_{n+1/2}).
\end{equation}
Using the implicit formulation of $f$, we obtain
\begin{equation}
\begin{aligned}
\bm f(t_n + \Delta t/2, \bm m_n, \bm m_{n+1}) =& (-\bm m_{n+1/2} \times \bm h_{n+1/2}) + \alpha (\bm m_{n+1/2} \times \Delta \bm m) \\
&- \bm m_{n+1/2} \times( \bm m_{n+1/2} \times \bm i_s),
\end{aligned}
\end{equation}
where
 $\bm m_{n+1/2} = (\bm m_n + \bm m_{n+1})/2$, $\bm h_{n+1/2} = \bm h(\bm m_{n+1/2})$, and $\Delta \bm m = (\bm m_{n+1} - \bm m_n)/ \Delta t$.
Since we are dealing with an implicit method, using the implicit definition of $d \bm m/dt$ from \eqref{eq_llg_implicit1} does not require any additional computation. In fact, using the implicit definition of $d\bm m/dt$ yields a more efficient method. This is due to the reduced complexity of the Jacobian of the system $\bm S_n$, which is used to solve for $\bm m_{n+1}$,
\begin{equation}
\label{eq_llg_implicit_system}
\bm S_n(\bm m) = \bm m - \bm m_n - \Delta t \cdot \bm f(t_n + \Delta t/2, \bm m_n, \bm m ).
\end{equation}

\noindent Having set up the system $\bm S_n$, it remains to derive its Jacobian with respect to $\bm m$.

\subsection{Derivation of the Jacobian}
First, we state that for $\bm a, \bm b \in \C^3$, the cross product of $a$ and $b$ can be restated in terms of matrix-vector multiplication
\begin{subequations}
\begin{equation}
\bm a \times \bm b = (a^{\times}) \bm b,
\end{equation}
where
\begin{equation}
a^{\times} = \begin{bmatrix}
0 & -a_3 & a_2 \\
a_3 & 0 & -a_1 \\
-a_2 & a_1 & 0
\end{bmatrix}.
\end{equation}
\end{subequations}

\noindent The Jacobian of a cross product of two functions $\bm f,\bm g: \C^n \to \C^3$ is given by \cite{Choroszucha2010}
\begin{equation}
J[\bm f \times \bm g] = f^{\times} J[\bm g] - g^{\times} J[\bm f] .
\end{equation}
 
 \noindent Using the above relation, one can derive the Jacobian as
  
 \begin{equation}
 \label{eq_jacobian} 
 \begin{aligned}
 J[\bm S_n](\bm m) = \ &I + \Delta t/2 \left( \bm m_{n+1/2}^{\times} J[\bm h] - \bm h_{n+1/2}^\times \right) \\
 & + \Delta t /2 \left[ \bm m_{n+1/2}^\times \bm i_s^\times - \left(\bm m_{n+1/2} \times \bm i_s \right)^\times \right] \\
 &- \alpha \bm m_n^\times,
 \end{aligned}
 \end{equation}
 
 \noindent where $\bm m_{n+1/2} = (\bm m+\bm m_n)/2$, $\bm h_{n+1/2} = \bm h(\bm m_{n+1/2})$, and
 \begin{equation}
 J[\bm h] = \frac{H_k}{M_s} \begin{bmatrix}
 n_x^2 & n_x n_y & n_x n_z \\
 n_y n_x & n_y^2 & n_y n_z \\
 n_z n_x & n_z n_y & n_z^2
 \end{bmatrix}
 - \begin{bmatrix}
 N_x & & \\
 & N_y & \\
 & & N_z
 \end{bmatrix},
 \end{equation}
 
 \noindent where $\bm n= (n_x, n_y, n_z)^T$ is the easy axis, and $\bm N = (N_x, N_y, N_z)^T$ are the demagnetization coefficients.
 To summarize, we will use the Jacobian in \eqref{eq_jacobian} to solve $\bm S_n$ in \eqref{eq_llg_implicit_system} for the magnetization value $\bm m_{n+1}$ using the Gauss-Newton algorithm.

 \section{Methods to solve the s-LLGS equation}\label{sec:methods}
In this section, we first obtain the Stratonovich SDE form of the s-LLGS equation. We then formulate the various numerical schemes for Stratonovich SDEs. Specifically, we focus on implicit midpoint, Heun, Euler-Heun, and a new method, namely the RK4-Heun. The accuracy of these methods is tested on an SDE with a known analytical solution. Finally, we perform numerical tests for these methods by applying them on the s-LLGS equation.

\subsection{Stratonovich SDE form of the s-LLGS equation}
  
\noindent A stochastic integral over a Wiener process $W(t): \R^+ \to \R$ \cite{Morters2010} is defined as
 \begin{equation}
 \int_a^b g(t ) dW(t) = \lim_{\Delta t \to 0} \sum_k g(\tau_k) (W(t_{k+1}) - W(t_k)),
 \end{equation}
 where $\{ t_k \}_k$ forms a partition of $[a,b]$, and $\tau_k \in [t_{k+1}, t_k]$.
 Contrary to the Riemann integral, different choices of $\tau_k$ yield different results of the integral.
 This is the case if $g$ is a function of both $t$ and the Wiener process $W(t)$.
 The most common choices are $\tau_k = t_k$ and $\tau_k = (t_{k+1} + t_k)/2$.
 The former yields the Ito calculus, while the latter results in the Stratonovich calculus.
 A widely used notation for the Stratonovich integral to distinguish it from the Ito integral is 
 \begin{equation}
 \int_a^b g(t) \circ dW(t),
 \end{equation}
which we will use in the following.
 Numerical schemes that attempt to approximate an SDE's solution have to be constructed for a specific calculus. In our case, this is the Stratonovich interpretation \cite{Gardiner1997}.\\
 
\noindent In a somewhat general form, we can write a Stratonovich SDE as
 \begin{equation}
 \label{eq_stochastic_def}
 dX_t = f(X_t, t) dt + g(X_t, t) \circ dW_t.
 \end{equation}
The s-LLGS equation in \eqref{eq_llg_explicit1} can be restated in the general Stratonovich SDE form given in \eqref{eq_stochastic_def}.
To this end, let $X(t) = \bm{m}_t$, and define
\begin{subequations}
\label{eq_fg_stochastic}
\begin{equation}
\label{eq_f_stochastic}
\begin{aligned}
f(\bm{m}_t, t) =& -\alpha' \ [ \bm{m}_t \times \bm{h} + \bm{m}_t \times (\bm{m}_t \times \bm{i}_s) \\
&+ \alpha (\bm{m}_t \times (\bm{m}_t \times \bm{h}) - \bm{m}_t \times \bm{i}_s )],
\end{aligned}
\end{equation}
 
\begin{equation}
\label{eq_g_stochastic}
\begin{aligned}
g(\bm{m}_t,t) =& - (\alpha' \nu) \ m_t^\times [ I + \alpha m_t^\times ], 
\end{aligned}
\end{equation}
\end{subequations}
where $\alpha' = 1/(1 + \alpha^2)$, and $\nu$ is the standard deviation of the Wiener process, which is used to model the thermal field fluctuations. 
Together \eqref{eq_stochastic_def} and \eqref{eq_fg_stochastic} define the form of the s-LLGS equation we will be working with throughout this paper. Further, we will deal with two modes of convergence of numerical approximations to solutions of SDEs
introduced in \cite{Kloeden_Platen}.
Given a solution $X_t$ of \eqref{eq_stochastic_def},
an approximation $\tilde X_t$ is said to converge to $X_t$ in the {\it strong sense with order} $\gamma > 0$ if
there is a $ C \in \R$ such that
\begin{equation}
\E \left( \left| \tilde X_t - X_t \right| \right) < C (\Delta t)^\gamma,
\end{equation}
holds for any discrete approximation $\tilde X_t$ with maximum step-size $\Delta t \>$\cite{Kloeden_Platen}.
On the other hand, an approximation $\tilde X_t$ is said to converge to $X_t$ in the {\it weak sense with order} $\gamma > 0$ if
there is a $ C \in \R$ such that
\begin{equation}
\left| \E ( p( \tilde X_t) ) -  \E (p(X_t)) \right| < C (\Delta t)^\gamma,
\end{equation}
holds for any polynomial $p$ and any discretization with maximum step size $\Delta t$.
Note that convergence in the strong sense is equivalent to convergence
for each realization of $W_t$ (i.e. path), while convergence in the weak sense 
merely implies that the statistical properties of the approximation converge to those of the solution.


\subsection{Numerical schemes for Stratonovich SDEs}
\begin{figure}[H]
  \begin{center}
    \includegraphics[scale=0.42]{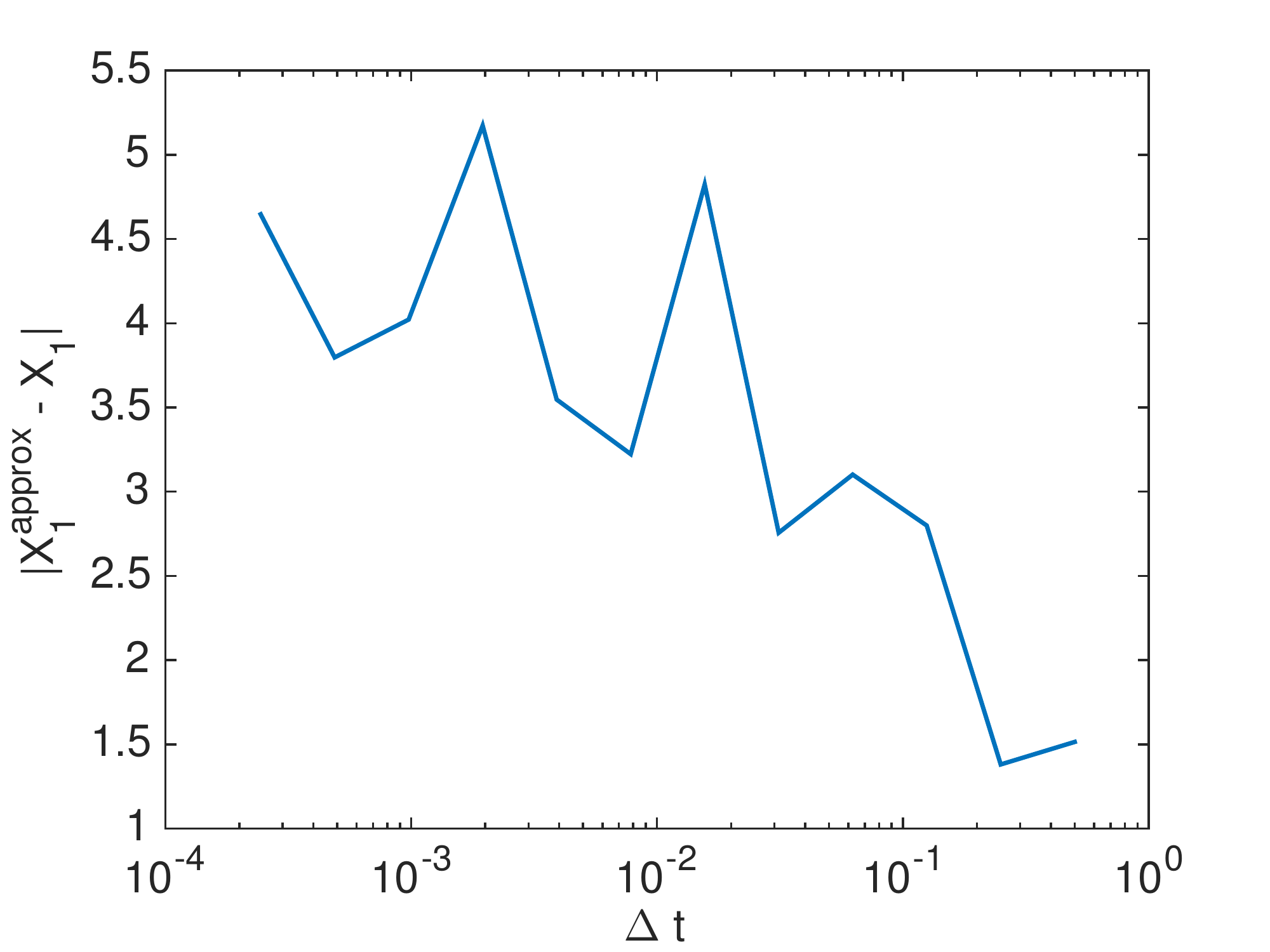}
  \end{center}
  \vspace*{-3ex}
  \caption{\footnotesize Average path-wise error of midpoint approximations using Adam's extrapolation 
  compared to the analytical solution of \eqref{eq_test_eq}.}
  \label{fig_adam_path}
\end{figure}

In order to numerically solve \eqref{eq_stochastic_def}, the following must be computed:
\begin{equation}
\label{eq_stochastic_solution}
X_t = \int_0^t f(X_t, t) dt + \int_0^t g(X_t,t) \circ dW(t) + X_0,
\end{equation}
where the first integral is a regular Riemann integral, and the second integral is a stochastic integral interpreted in the Stratonovich sense. In our case, we are dealing with three-dimensional (3D) vector integrals, and $W(t)$ is a 3D Wiener process. Below, we briefly outline the methods that we focus on in this work.
\begin{enumerate}

\item \textit{Euler-Heun}\\
Arguably the simplest method that converges to the Stratonovich solution is the Euler-Heun method defined by
\begin{subequations}
\begin{equation}
X_{n+1} = X_n+ f(X_n, t_n) \Delta t + \frac{1}{2} \left[ g( \tilde X_{n+1}, t_{n+1}) + g(X_n, t_n)\right] \eta_n,
\end{equation}
\begin{equation}
\tilde X_{n+1} = X_n + g(X_n, t_n)\eta_n,
\end{equation}
where $\eta_n = \sqrt{\Delta t} \xi_n$, $\Delta t \in \R^+$, and $\xi_n \sim \mathcal{N}(0,1)$.
\end{subequations}
\item \textit{Heun}\\
The Heun method is given by
\begin{equation}
X_{n+1} = X_n + \frac{1}{2} \left[ f(\tilde X_{n+1}, t_{n+1}) + f(X_n, t_n) \right] \Delta t+  \frac{1}{2}\left[ g( \tilde X_{n+1}, t_{n+1}) + g(X_n, t_n)\right] \eta_n,
\end{equation}
where
\begin{equation}
\begin{aligned}
\tilde X_{n+1} = X_n + f(X_n, t_n) \Delta t +  g(X_n, t_n) \eta_n.
\end{aligned}
\end{equation}
\item \textit{Implicit midpoint}\\
The implicit midpoint rule is defined by
\begin{equation}
\begin{aligned}
X_{n+1} = X_n + f(X_{n+1/2}, t_n+\Delta t/2) \Delta t + g(X_{n+1/2}, t_n+ \Delta t/2) \eta_n, \\
\end{aligned}
\end{equation}
where $X_{n+1/2} = (X_{n+1} + X_n) /2$.
Note that this defines an implicit system just as in the deterministic case, and we can use the methods derived for the deterministic case to solve it.
We use the Euler step
\begin{equation}
X^* = X_n + f(X_n, t_n) \Delta t + g(X_n, t_n) \eta_n,
\end{equation}
as the initial guess for the Gauss-Newton algorithm, to find $X_{n+1}$.\\

\end{enumerate}
\vspace*{-3ex}
\noindent All of the above methods are known to converge to the Stratonovich solution 
with a strong order $1/2$, and a weak order $1$ \cite{Kloeden_Platen,Milstein2002}.
Compared to numerical methods for deterministic DEs, 
this order of convergence is low.
However, stochastic higher-order methods generally require the approximation of iterated stochastic integrals,
which is a detailed and time-intensive procedure \cite{Gaines_1993}.
If an SDE has a special structure, 
like an additive or commutative noise term,
this calculation can be simplified.
Unfortunately, the noise term of the s-LLGS equation is
multiplicative and non-commutative.\\
\noindent Further, it is important to note that semi-implicit numerical methods based on extrapolation, which are used to handle deterministic DEs, generally do not converge to solutions of SDEs.
The reason for this is that extrapolations from the past destroy the Markov property (independent increments) of the Wiener process over which we integrate. 
As seen in Figure \ref{fig_adam_path}, a semi-implicit midpoint scheme using Adam's extrapolation \cite{Serpico2001}, where $X_{n+1/2} = \frac{1}{2}(3X_n-X_{n-1})$, does not converge to the solution of the test SDE,
\begin{equation}
\label{eq_test_eq}
dX_t = X_t dt + X_t \circ dW_t,
\end{equation}
with solution
\begin{equation}
\label{eq_test_sol}
X_t = e^{t+W_t}.
\end{equation}

\vspace*{-3ex}

\subsection{A method for small noise}

In many stochastic models corresponding
to physical systems, the noise term is usually much smaller than
the drift term.
Indeed, while the stochastic term plays a critical role
in the s-LLGS equation, it is frequently one to two orders
of magnitude smaller than the deterministic term,
if $T$ is close to room temperature.
This means a higher order approximation of the 
drift term alone could improve the accuracy 
of the entire approximating process,
when the absolute error is dominated by
the error in the deterministic term.
This idea has been investigated for Ito SDEs in \cite{Mil_1997, Ro_2013}.
Here, we propose a scheme for Stratonovich SDEs, which
we are going to refer to as RK4-Heun.
It is defined by

\begin{equation}
\begin{aligned}
X_{n+1} &= X_n + D \Delta t + S \eta_n, \\
D &= (d_1 + 2d_2 + 2d_3 + d_4)/6, \\
S &= (s_1 + s_2)/2,\\
s_1 &= g(X_n,t_n),\\
s_2 &= g(X_n + f(X_n, t_n) \Delta t + s_1 \eta_n), \\
d_1 &= f(X_n,t_n),\\
d_2 &= f(X_n + (d_1 + s_1)/2, t_n + \Delta t/2), \\
d_3 &= f(X_n + (d_2 + s_1)/2, t_n + \Delta t/2), \\
d_4 &= f(X_n + d_3 + s_1, t_n + \Delta t).
\end{aligned}
\end{equation}

\noindent Importantly, the RK4 stages are only used to approximate
the deterministic part of the equation.
Therefore, convergence to the Stratonovich solution
is maintained.
Further, we made the observation that making the additional update
\begin{equation}
\begin{aligned}
X_{n+1}' &= X_n + D \Delta t + S' \eta_n, \\
S' &= (s_1 + s_2')/2, \\
s_2' &= g(X_{n+1}, t_{n+1}), \\
\end{aligned}
\end{equation}
after computing the above, leads to a significant
increase in accuracy for the test equation \eqref{eq_test_eq}.
The additional update includes the higher order approximation
of the deterministic term in the Heun step of the 
stochastic term.
As a consequence, it is not surprising that this
leads to a better approximation.
However, we have not yet investigated 
analytical justifications of this update.
Future work will look into that
 and the possible limitations of this effect.

\subsection{Numerical tests using a general SDE}

To compare the accuracy of the various methods discussed above, we consider
a modified version of the test SDE \eqref{eq_test_eq}, which can be solved analytically,
\begin{equation}
\label{eq_test_gen}
dX_t = a X_t dt + b X_t \circ dW_t,
\end{equation}
where $a, b \in \R$.
A Stratonovich solution to \eqref{eq_test_gen} is given by
\begin{equation}
\label{eq_integral_test_eq}
X_t = X_0 + \int_0^t X_s ds + \int_0^t X_s \circ dW_s = e^{at+b W_t},
\end{equation}
where $X_0 = 1$.
We can now compare the results of different numerical schemes to this analytical solution. 

\noindent The implicit midpoint method for \eqref{eq_test_eq} is defined by
\begin{equation}
X_{t+1} = X_t + a \frac{1}{2} ( X_{t+1} + X_t) \Delta t + b \frac{1}{2}(X_{t+1} + X_t) \eta_t.
\end{equation}
This can be solved for $X_{t+1}$ analytically, according to 
\begin{equation}
\begin{aligned}
\label{eq_test_implicit}
X_{t+1} = X_t \left( \frac{ 2 + c}{2 - c} \right), \\
c = a \Delta t + b \eta_t. 
\end{aligned}
\end{equation}

\begin{figure}[H]
  \centering
  \begin{subfigure}{.45\linewidth}
  \centering
    \includegraphics[scale=0.4]{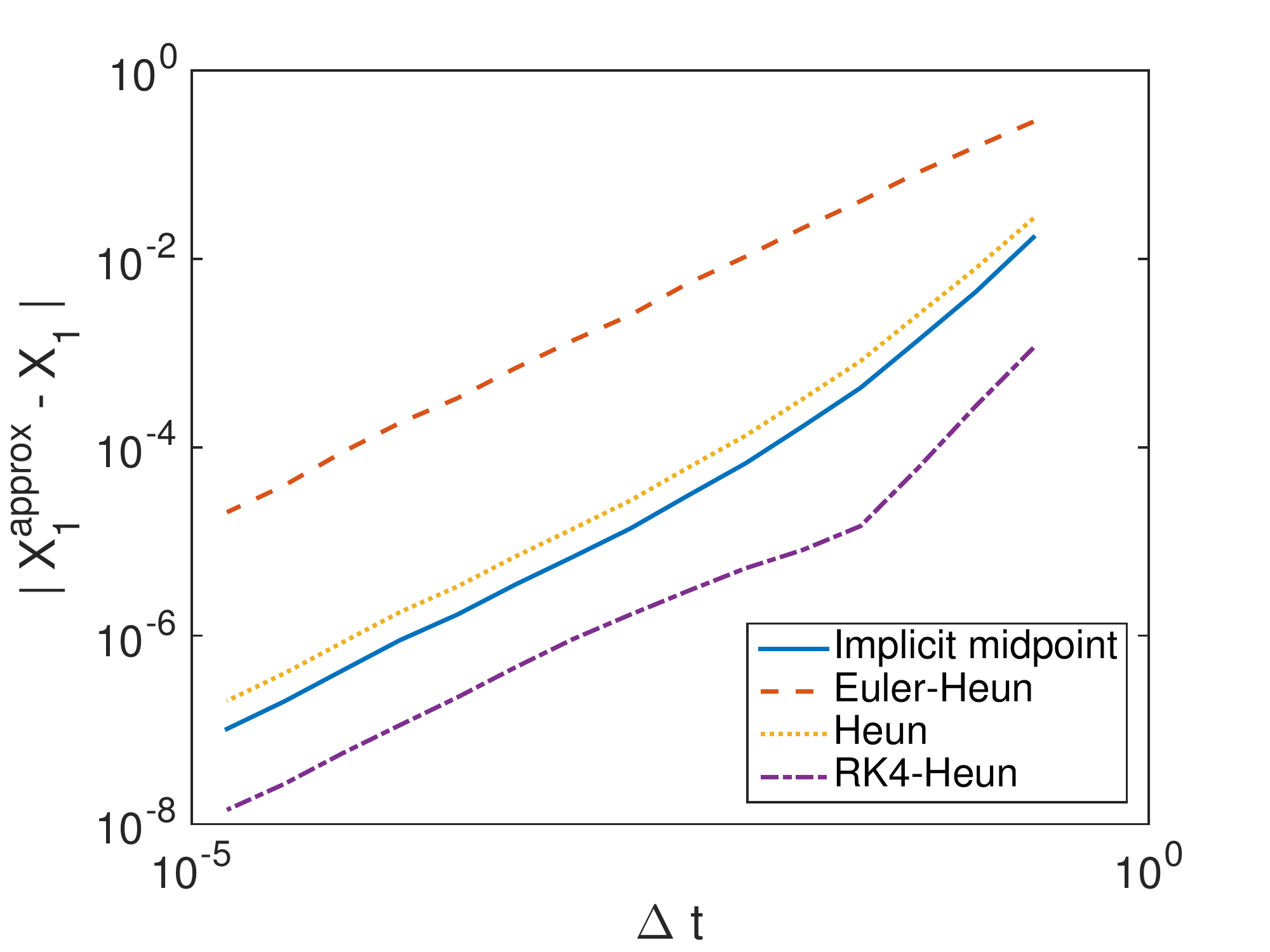} 
   \vspace{-3ex}
  \caption{$a = 1, b = 0.1$}
  \label{fig_path_err_1}
  \end{subfigure}
  \qquad
  \begin{subfigure}{.45 \linewidth}
  \centering
  \includegraphics[scale=.4]{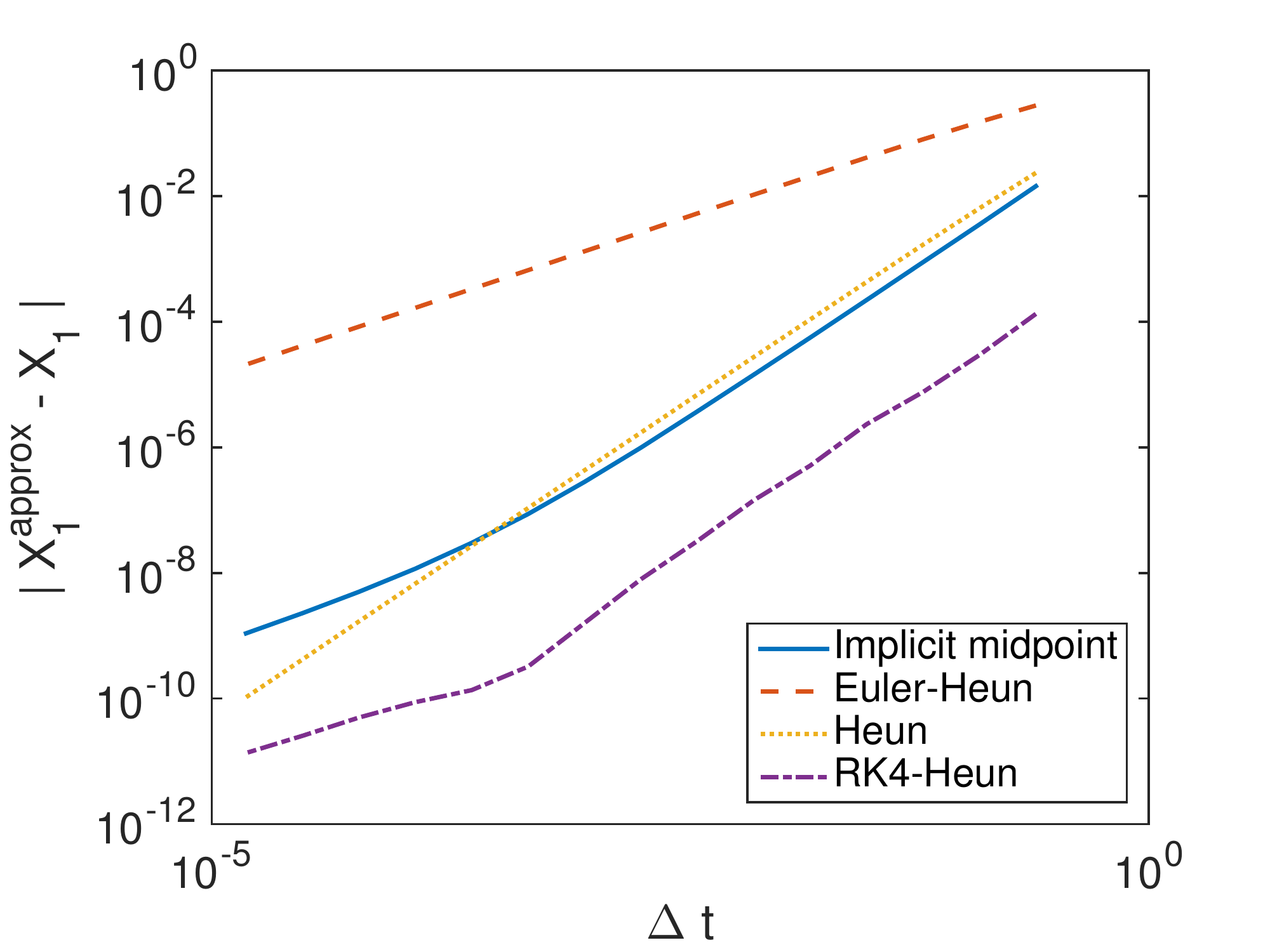}
  \vspace{-3ex}
  \caption{$a = 1, b = 0.01$}
  \label{fig_path_err_2}
  \end{subfigure}
 \vspace{-1ex}
  \caption{\footnotesize Average path-wise error of the numerical approximations
  corresponding to equation \eqref{eq_test_gen} with two different choices
  of $a$ and $b$. }
  \label{fig_path_err}
\end{figure}

\noindent The experimental results for the path-wise error to the analytical solution, $X_{1}$, are shown in Figure \ref{fig_path_err_1} and \ref{fig_path_err_2} for two different choices of $b$ in \eqref{eq_test_gen}.
The data shows that the RK4-Heun scheme is more accurate
than all other methods for all step sizes considered.
Interestingly, the empirical order of convergence, which is the slope of the error function
in Figure \ref{fig_path_err}, is only constant for the Euler-Heun method.
All other methods start out with a faster order (steeper slope) for larger step sizes.
This is due to the fact that the
deterministic part of the equation, and hence the error in its approximation,
 dominates the stochastic part.
 Therefore, the slope for big step sizes is more like the deterministic second
 order of convergence
 for the Heun and implicit midpoint schemes.
We see this effect for smaller step sizes as the size $b$ of the stochastic part is decreased in 
Figure~\ref{fig_path_err_2}.
\vspace{-2ex}
\vspace{-3ex}

\subsection{Numerical tests using the s-LLGS equation}

\noindent Now, we are going to test the properties of these methods on the s-LLGS equation, defined by \eqref{eq_f_stochastic} and \eqref{eq_g_stochastic}.
It is important to note that the norm of the magnetization vector $\bm{m}$ is preserved if the equation is interpreted in the Stratonovich sense.
This follows from the fact that in the Stratonovich calculus,
\[
d( \| m_t \|^2 ) = 2 \bm{m}_t \cdot d\bm{m}_t = 0,
\]
whereas in the Ito calculus, Ito's Lemma gives
\vspace{-2ex}
\[
d( \|m_t\|^2 ) = 2 \bm{m}_t \cdot d\bm{m}_t + 
\sum_{i,j = 1}^3 \frac{ \partial^2 \| \bm m\|^2}{\partial m_i \partial m_j}(\bm{m}_t) ~ \left( dm_{i,t} \cdot dm_{j,t} \right),
\]
which is, in general, not zero.
Explicit schemes like Euler-Heun or Heun, while solving the Stratonovich equation, do not preserve the norm.
Indeed, one has to take a very small step size $\Delta t$ so that the norm of the magnetization does not blow up.
On the other hand, the implicit midpoint method preserves the norm.
The contrasting behavior of the magnetization norm obtained using different stochastic calculi is highlighted in Figures 3a and 3b. Indeed the midpoint method that converges to the Stratonovich solution preserves the norm while the Ito solution does not.

\vspace{-2ex}
\begin{figure}[H]
\centering
\begin{minipage}{.5\textwidth}
  \centering
  \includegraphics[scale=0.39]{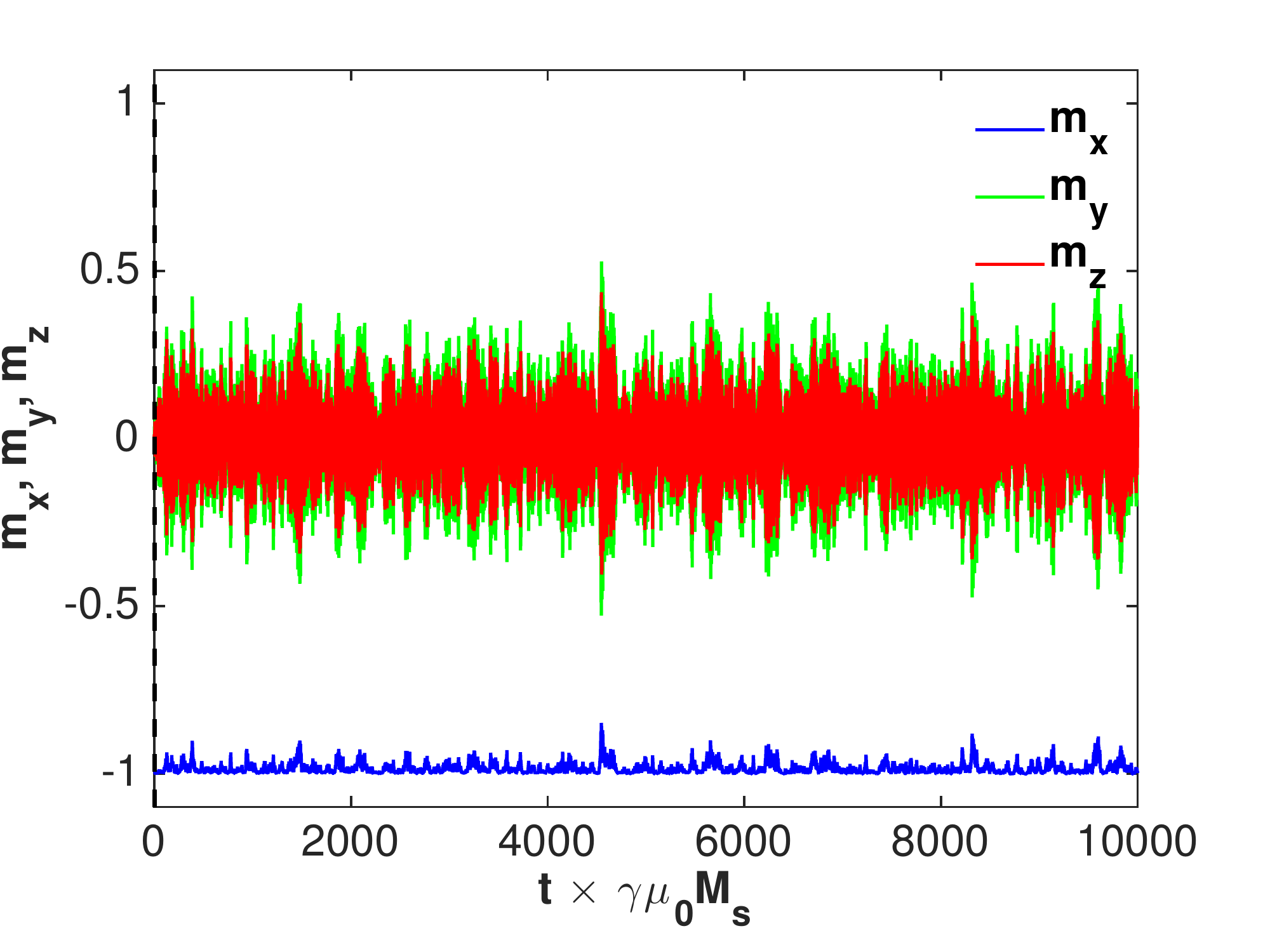}
  {(a)}
  \label{fig:test1}
\end{minipage}%
\begin{minipage}{.5\textwidth}
  \centering
  \vspace*{-2ex}
  \includegraphics[scale=0.39]{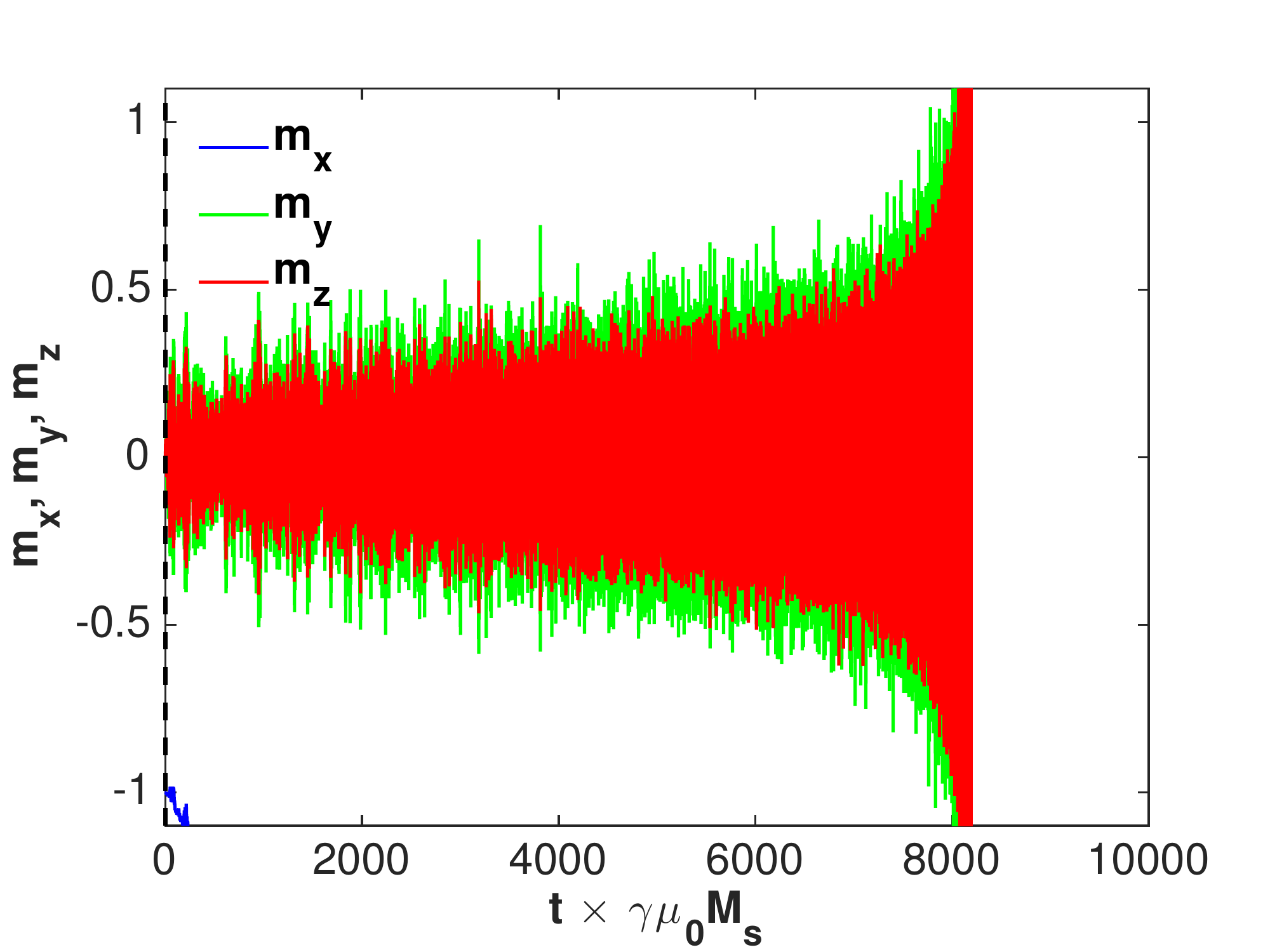}
  (b)
  \label{fig:test2}
\end{minipage}
\vspace{-2ex}
\caption{\footnotesize Time evolution of magnetization with (a) implicit midpoint converging to the Stratonovich solution and (b) Heun-Euler converging to the Ito solution. Conditions are zero spin current (only noise) and energy barrier $U = 10K_BT$. The magnetization norm is preserved with the former, but blows up with the latter. }
\end{figure}

\begin{figure}[H]
\centering
\begin{subfigure}{.45\textwidth}
  \centering
  \includegraphics[scale=0.4]{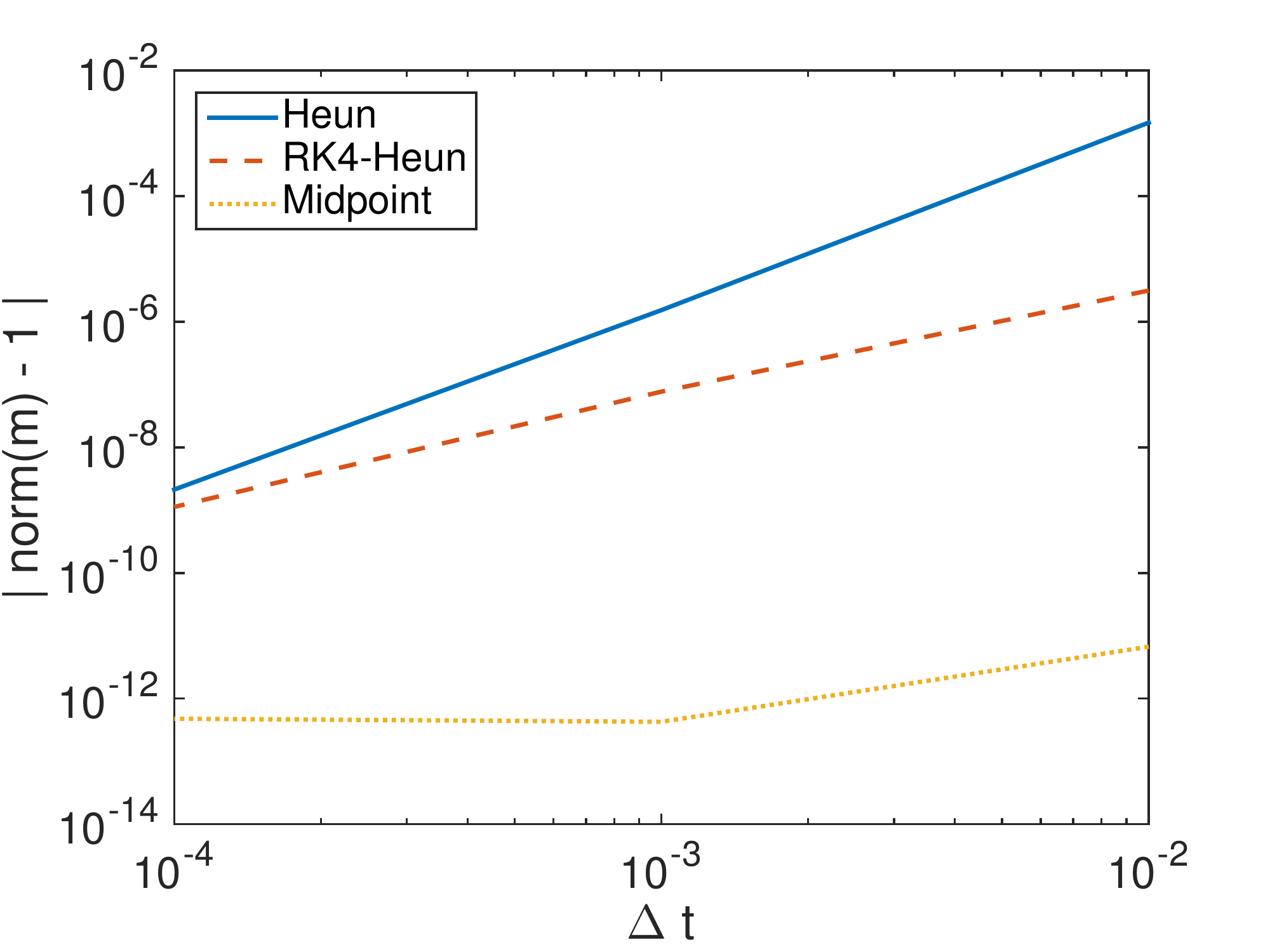} 
  \caption{\footnotesize Norm of magnetization vector in the course of several simulations with different $\Delta t$, 
  calculated with three numerical schemes.
  Here, $1$ ns $\sim 66$ time units.}
  \label{fig_norm_comparison}
  \end{subfigure}
  \qquad
\begin{subfigure}{.45\textwidth}
  \centering
  \includegraphics[scale=0.4]{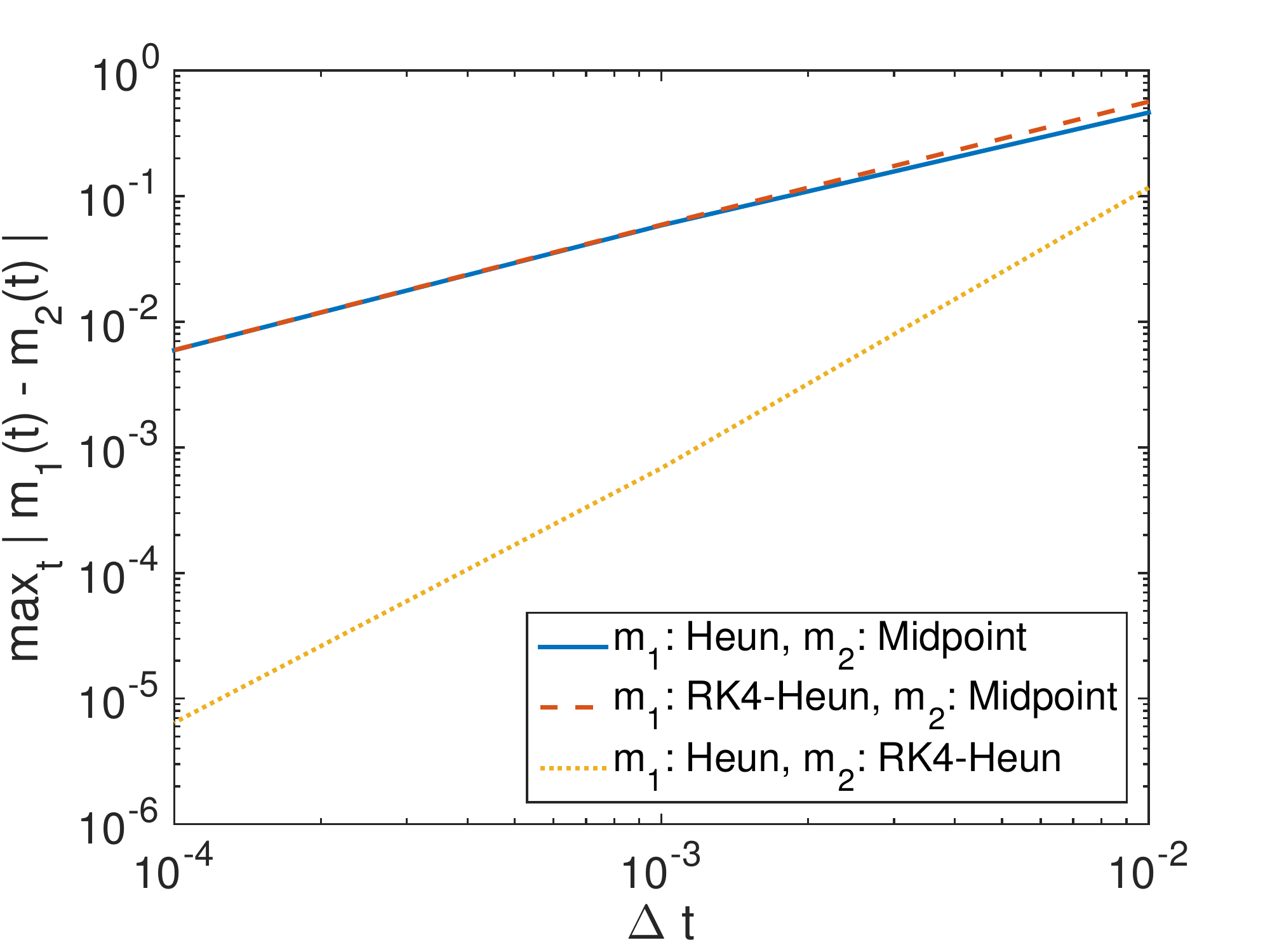}
  \caption{\footnotesize Maximum norm of the difference of magnetization vectors as calculated with the implicit midpoint method, Heun, and RK4-Heun scheme. }
  \label{fig_magnetization_difference}
\end{subfigure}
\caption{\footnotesize Results of using the Heun, RK4-Heun, and implicit midpoint scheme on the s-LLGS equation.}
\label{fig_LLG_bench}
\end{figure}

\noindent The deviation of the magnetization norm from unity for different numerical techniques is depicted in Figure \ref{fig_norm_comparison}.
The error tolerance for the Gauss-Newton algorithm in the implicit midpoint method 
was $10^{-12}$.
For this reason, the norm is only preserved up to this order of magnitude. 
In practice, decreasing the error tolerance further only slows down the algorithm
without significant benefit. In Figure \ref{fig_magnetization_difference}, we show the maximum
 norm of the difference of the magnetization vectors, calculated by the Heun, implicit midpoint, and RK4-Heun methods. The difference is quite pronounced at larger time steps as expected, and diminishes as the time step is reduced.
 We see that the explicit methods converge to the same path quickly,
while the difference to the implicit scheme decreases at a slower rate.\\

\vspace{-4ex}

\begin{figure}[H]
    \includegraphics[scale=0.47]{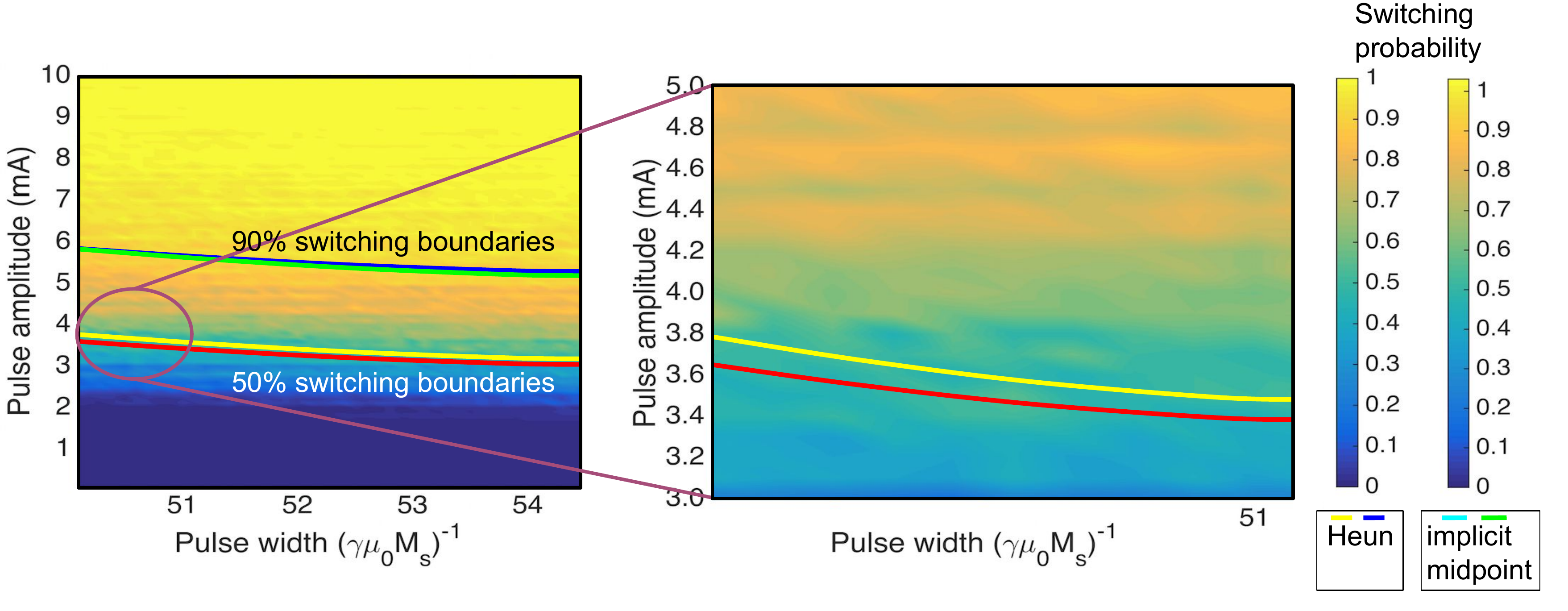}
    \vspace{-4ex}
  \caption{\footnotesize Contour plot for switching probability as a function of spin current amplitude and duration. 50$\%$ and 90$\%$ switching boundaries are depicted for implicit midpoint and explicit Heun schemes.}
  \label{3Dcontour}
  \vspace{0ex}
\end{figure}

\noindent It would be interesting to see the discrepancy in the switching boundaries obtained from the Heun and implicit midpoint schemes. The 50$\%$ switching boundary is a useful metric since it is commonly used to benchmark solvers. To this end, we construct a contour plot (Figure \ref{3Dcontour}) by sweeping over spin current durations and amplitudes to ascertain the 50$\%$ and 90$\%$ switching boundaries from the two methods. The maximum discrepancy in the switching boundaries is observed at the 50$\%$ switching probability. 
However, the difference in the switching probability between the two methods diminishes in the high current, long time-scale regimes for the chosen time step of numerical integration.

\vspace{-4ex}

\section{SPICE simulation of macrospin dynamics}

Circuit-level simulations using nanomagnets are often conducted using SPICE-based models, which represent the nanomagnet using basic circuital elements such as resistors, capacitors, and current sources. 
It is, therefore, important to benchmark the accuracy of SPICE numerical solvers with respect to the implicit midpoint method to solve the s-LLGS equation.  SPICE uses methods such as Euler, Gear, trapezoidal, or a combination of these, to numerically integrate DEs. Although the trapezoidal method is compatible with the Stratonovich stochastic calculus, other numerical methods in SPICE require that the s-LLGS equation be transformed into the equivalent Ito representation for correct convergence. While there are many versions of SPICE available \cite{ngspice2011,xyce2014,smartspice2010}, possibly with varying minimum time steps and specific algorithms to solve equations, we use NGSPICE for our simulatons in this paper. Figure~\ref{fig:nanomagnet_schem} shows the SPICE circuit implementation of a nanomagnet subject to thermal effects and spin torque, in one particular direction (x, y or z). The other two magnetization directions are implemented similarly. The
magnetization $\bm{m} = \bm{M}/M_s$ is modeled as the node voltage of a capacitor, and the effective field inside the nanomagnet is represented as a dependent current source.  Detailed circuit derivations can be found in prior works \cite{Sasikanth2012} and \cite{Phillip}. The s-LLGS equation was implemented in SPICE in both its cartesian and spherical coordinates form, and the results from these implementations are discussed below. 

\subsection{SPICE implementation of the s-LLGS equation in cartesian form}
\begin{figure}[H]
    \centering
    \begin{subfigure}[t]{0.5\textwidth}
        \centering
        \includegraphics[scale=0.47]{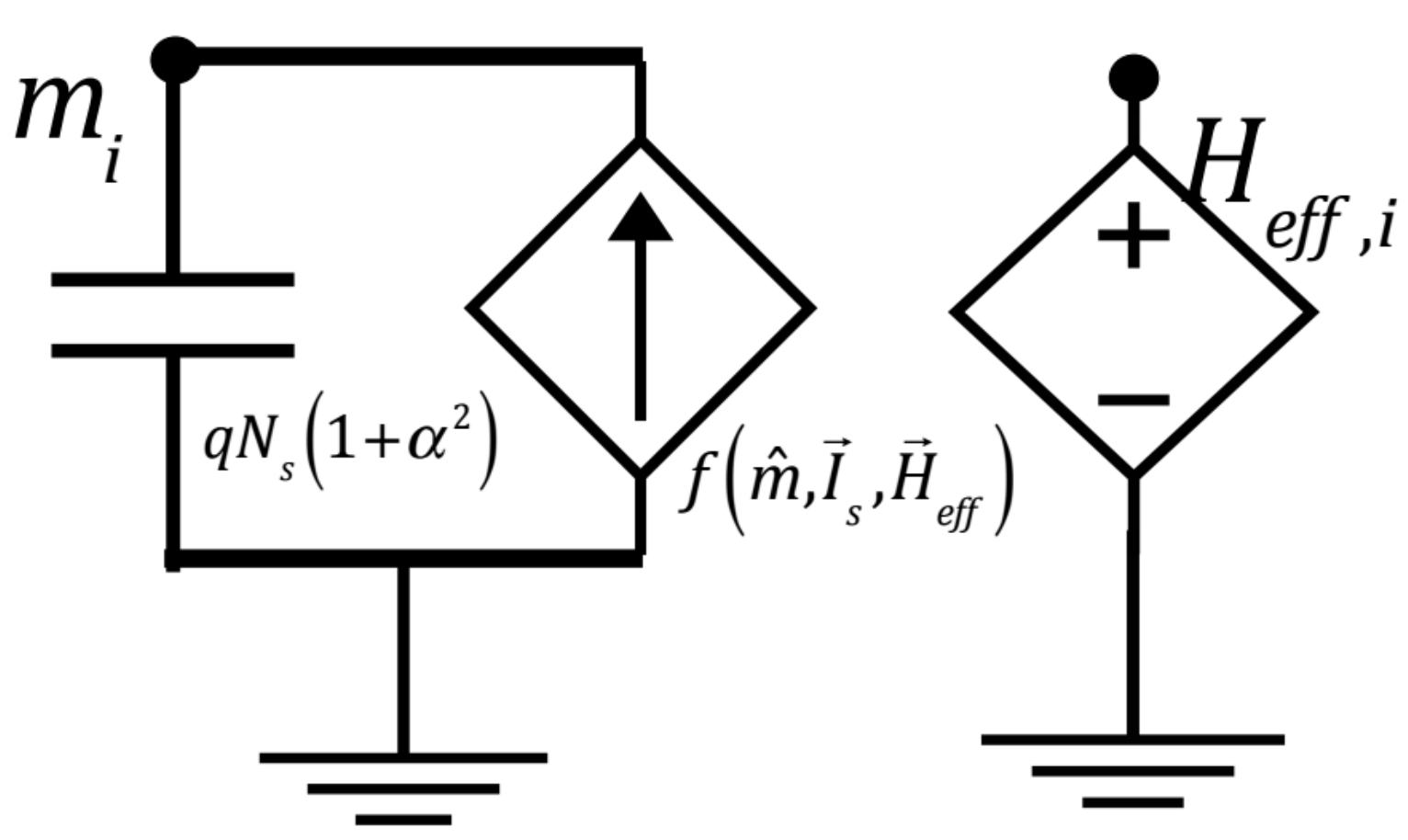}
        \caption{\footnotesize Schematic of nanomagnet used in the circuit\\ simulations.}
        \label{fig:nanomagnet_schem}
    \end{subfigure}%
    ~ 
    \begin{subfigure}[t]{0.5\textwidth}
        \centering
        \includegraphics[scale=0.33]{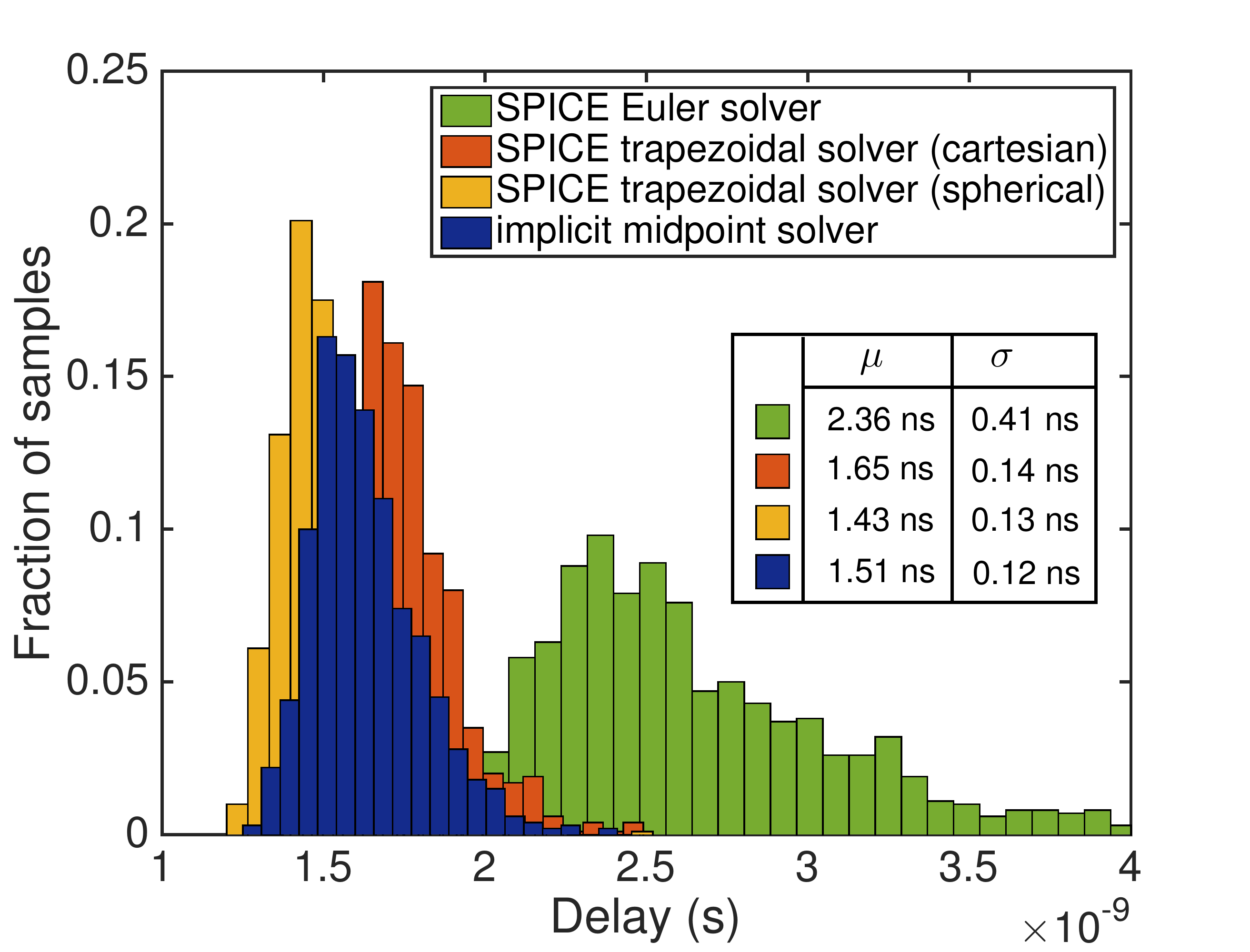}
        \caption{\footnotesize PDFs of reversal delays of a nanomagnet using implicit midpoint, SPICE trapezoidal solver (cartesian and spherical s-LLGS), and SPICE backward-Euler solver.}
        \label{fig:spice_vs_matlab}
    \end{subfigure}
    \vspace*{-1ex}
    \caption{\footnotesize The simulations were performed for an in-plane magnet using identical dimensions $40\times 40\times 1$ nm$^3$, parameters $M_s = 1.11e6$ A/m, $H_k = 1.11e5$ A/m, $\alpha = 0.01$, time step $1$ ps, and spin current $0.16$ mA.}
\end{figure}

\begin{figure}[H]
\centering
\begin{minipage}{.5\textwidth}
  \centering
  \includegraphics[scale=0.4]{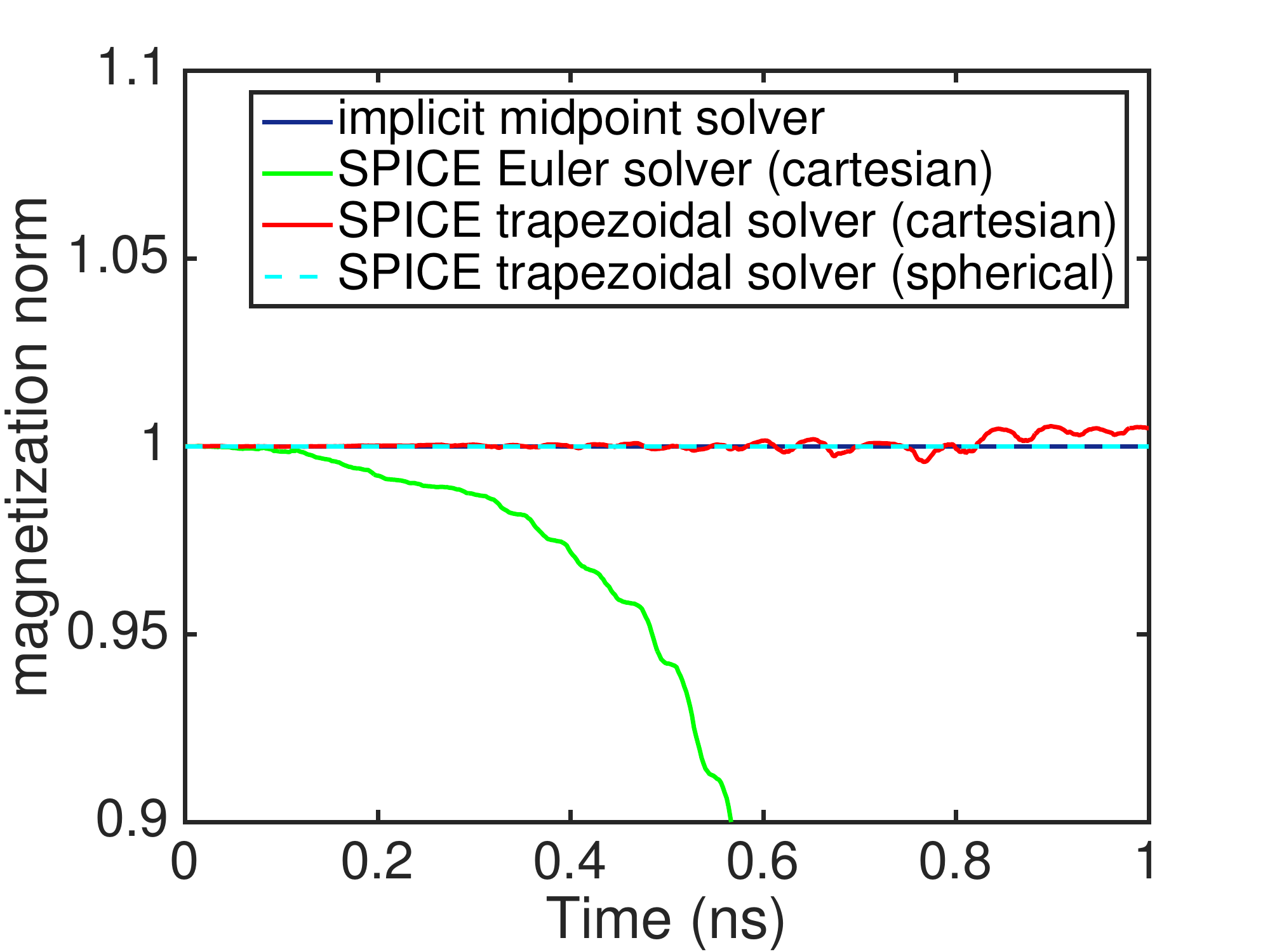} 
  \caption{\footnotesize Norm of the magnetization vector in the course of a simulation, calculated with SPICE and implicit mid-point solvers. The parameters used are the same as for Figure~\ref{fig:spice_vs_matlab}. }
  \label{SPICE_norm}
  \end{minipage}%
 ~
  \hspace*{2ex}
\begin{minipage}{.5\textwidth}
  \centering
  \vspace{2ex}
  \includegraphics[scale=0.42]{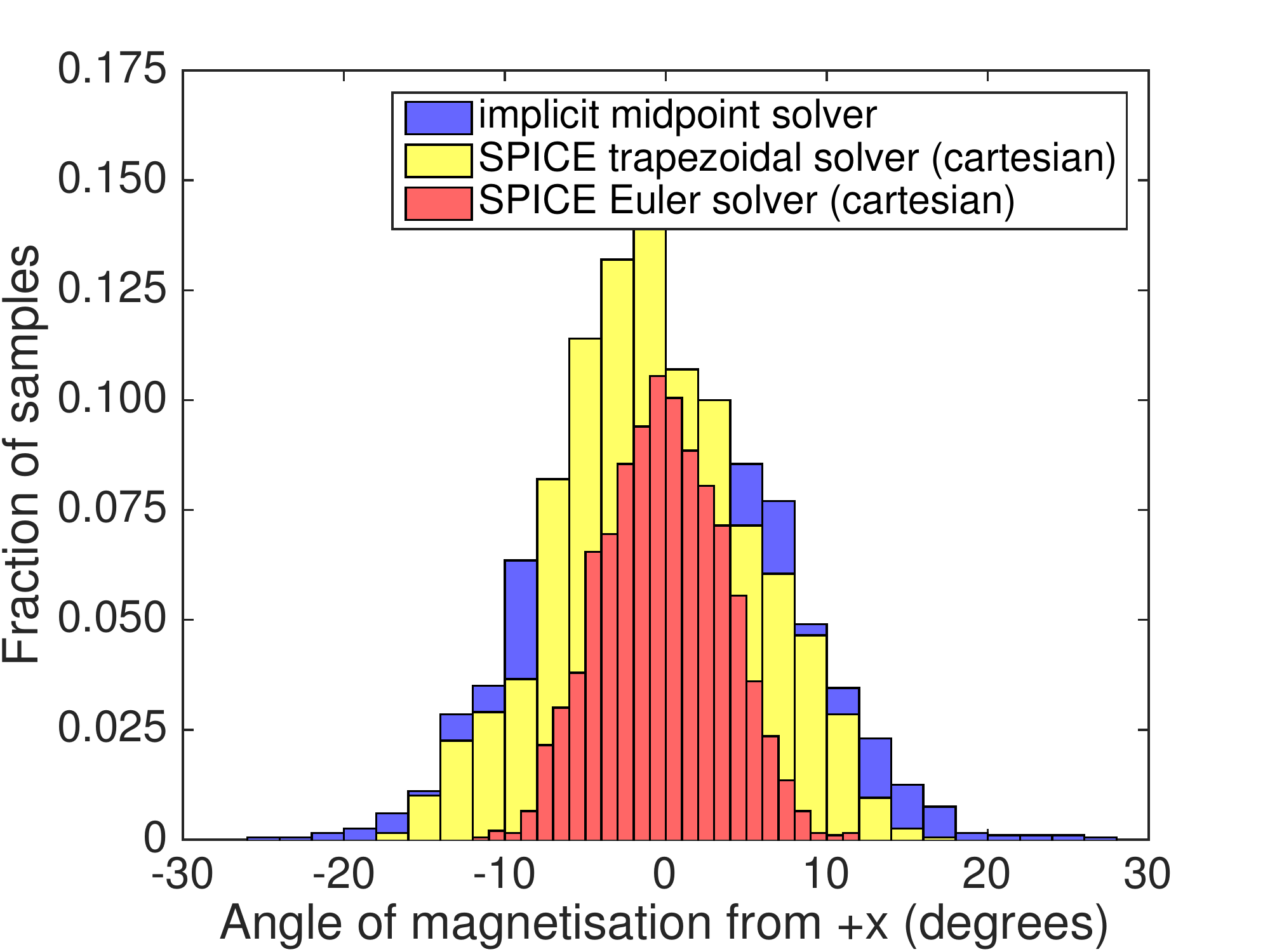}
  \caption{\footnotesize Initial angle distribution of the nanomagnet after it has been subjected to only thermal noise for 1 ns, from SPICE and our implicit midpoint solver. \\}
  \label{initial_angle}
\end{minipage}
\end{figure}

Large-scale Monte Carlo simulations were conducted using the circuit representation in Figure~\ref{fig:nanomagnet_schem} to generate the probability density function (PDF) plots of the magnetization reversal delays of an in-plane nanomagnet. The time step of integration is taken to be 1 ps, reducing which, leads to convergence issues in the SPICE nodal analysis. From the results shown in Figure~\ref{fig:spice_vs_matlab}, we find that SPICE solvers working with the cartesian form of the s-LLGS equation overestimate the mean delay of magnetization reversal as compared to the implicit midpoint method. 
In addition, unlike the implicit midpoint method, SPICE solvers are unable to preserve the magnetization norm with the cartesian s-LLGS eqauation, as shown in Figure \ref{SPICE_norm}. We also highlight the differences in the initial angle distribution obtained from various numerical schemes in Figure \ref{initial_angle}. In this figure, the macrospin is subject to thermal noise for a period of 1 ns and no external magnetic field or spin torque is applied. The s-LLGS equation conserves magnetization norm and has a Boltzmann initial angle distribution only if the Stratonovich stochastic calculus with the midpoint prescription is used. For other stochastic discretization schemes, none of these properties are ensured and a carefully chosen drift term has to added to recover the validity of these properties when other stochastic calculi are used \cite{roma2014numerical}. As expected, the discrepancy in results is exacerbated for Euler method due to the fact that it does not solve for the correct Stratonovich SDE equation and will require an alternative Ito representation of the s-LLGS equation. Whereas the trapezoidal method does converge to the Stratonovich solution for $\Delta t \rightarrow 0$, but it is limited by the minimum time step of the DE solver in SPICE. Hence we see a 200 ps difference in the mean reversal delay obtained with implicit midpoint and the trapezoidal solver of SPICE (with cartesian s-LLGS equation). It must be noted that these significant errors in magnetization reversal computed from SPICE will ultimately lead to inaccurate estimates of circuit-level performance metrics such as error-rate, latency, and energy dissipation of complex magneto-logic networks. \\

\vspace*{-4ex}

\vspace*{-2ex}

\subsection{SPICE implementation of the s-LLGS equation in spherical form}

To mitigate the aforementioned errors in the SPICE-based solvers, we look at the implementation of the s-LLGS equation in spherical coordinates form. The spherical coordinates of the unit magnetization vector $\mathbf{m}$ at any position can be written as $(\theta, \phi)$, where
$\theta$ is measured from the positive direction of $z$-axis, and $\phi$ is measured 
counterclockwise from the positive direction of $x$-axis. Further, the local orthogonal unit vectors in the directions of increasing $\theta$, and $\phi$ respectively are given by
\begin{equation}
\boldsymbol{\hat{\theta}} = [\cos\theta \cos\phi, \cos\theta \sin\phi, -\sin\theta]; \quad 
\boldsymbol{\hat{\phi}} = [-\sin\phi, \cos\phi, 0].
\end{equation}

\noindent The s-LLGS equation in spherical coordinates can be expressed in terms of the $\theta$ and $\phi$ components of the effective field ($h_{\theta}, h_{\phi}$), and spin current ($i_{s\theta}, i_{s\phi}$) as
\begin{equation}
\begin{bmatrix}
\frac{d\theta}{dt}\\ \sin\theta \cdot \frac{d\phi}{dt}
\end{bmatrix}
= \frac{1}{1+\alpha^2}\begin{bmatrix}	
h_{\phi} + i_{s\theta} + \alpha h_{\theta} - \alpha i_{s\phi}\\ 
i_{s\phi} - h_{\theta} + \alpha h_{\phi} + \alpha i_{s\theta}
\end{bmatrix},
\label{eq:LLG_sph}
\end{equation}

\begin{wrapfigure}{l}{0.52\textwidth}
\vspace{-5ex}
        \includegraphics[scale=0.31]{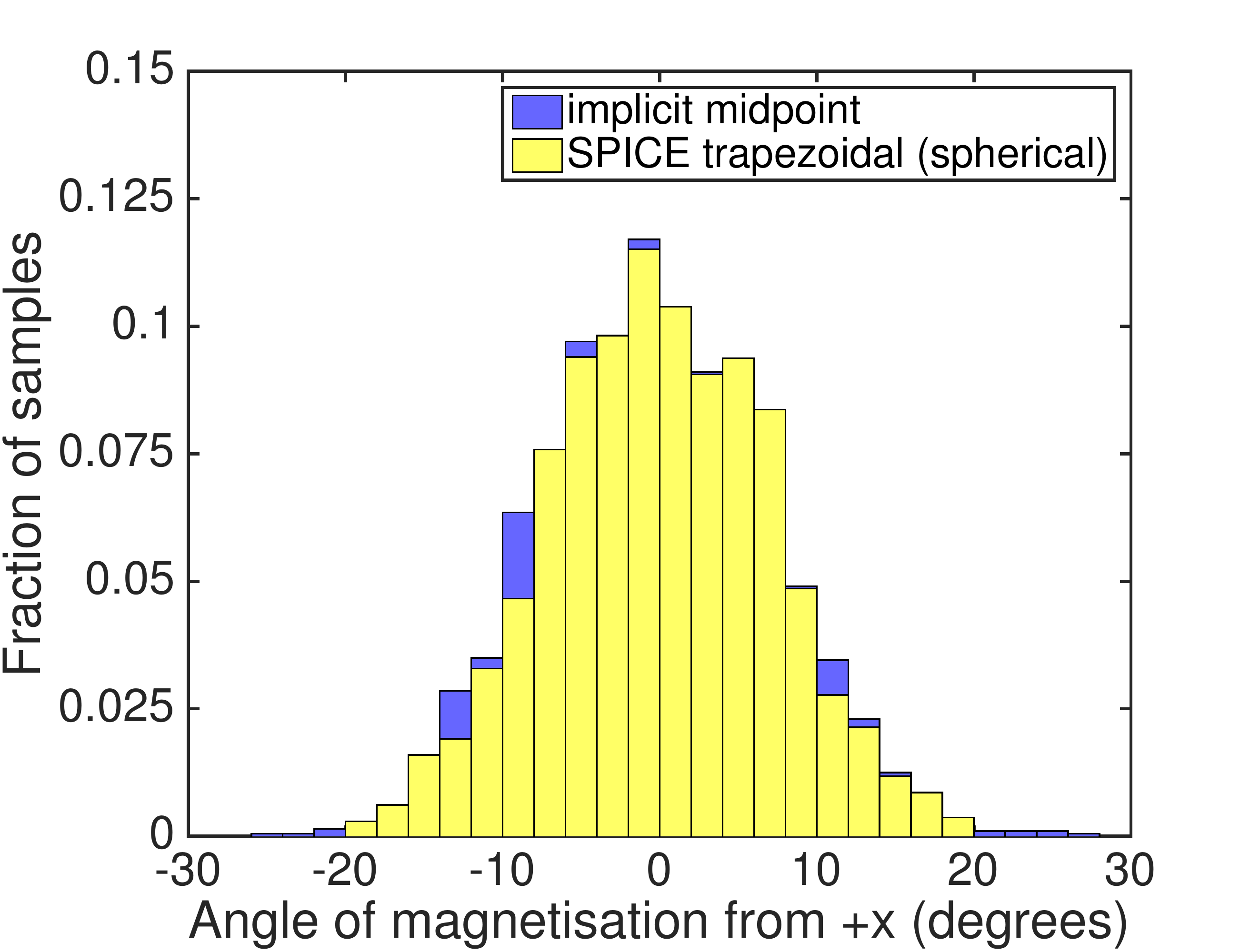}
        \caption{\footnotesize Initial angle distribution of the nanomagnet after it has been subjected to only thermal noise for 1 ns, from SPICE trapezoidal solver (spherical s-LLGS) and our implicit midpoint solver}
        \label{spherical_init}
    \vspace*{-1ex}
\end{wrapfigure}

\noindent where $(i_s/h)_{\theta} = \mathbf{(i_s/h)}\cdot\boldsymbol{\hat{\theta}}$, $(i_s/h)_{\phi} = \mathbf{(i_s/h)}\cdot\boldsymbol{\hat{\phi}}$, and $\cdot$ denotes dot product. The solution ($\theta$, $\phi$) of (\ref{eq:LLG_sph}) can be reverted back to cartesian coordinates by using $\mathbf{m} = [\sin\theta\cos\phi, \sin\theta\sin\phi, \cos\theta]$. Though a change of coordinate system does not change the solution of the system being solved, there are some subtle differences in the numerical implementation of the cartesian and spherical forms. An apparent difference is the need to solve for only two variables $\theta$ and $\phi$ in spherical coordinates, compared to three variables $m_x, m_y$ and $m_z$ in cartesian. This makes implementing s-LLGS equation in spherical coordinates less expensive, especially when using implicit methods, where the overhead cost of calculating the $\theta$ and $\phi$ components (few multiplications) is way less than solving the implicit step (which incurs fixed point iteration). This also forces the norm to be preserved to unity automatically as shown in Figure \ref{SPICE_norm}. Importantly, we can also see from Figures \ref{fig:spice_vs_matlab} and \ref{spherical_init} that the deviation in the delay distribution and initial angle distribution from the implicit midpoint solver is minimized for the case of the spherical s-LLGS implementation.

\begin{figure}[H]
    \begin{subfigure}[t]{0.33\textwidth}
 \includegraphics[width=0.95\textwidth]{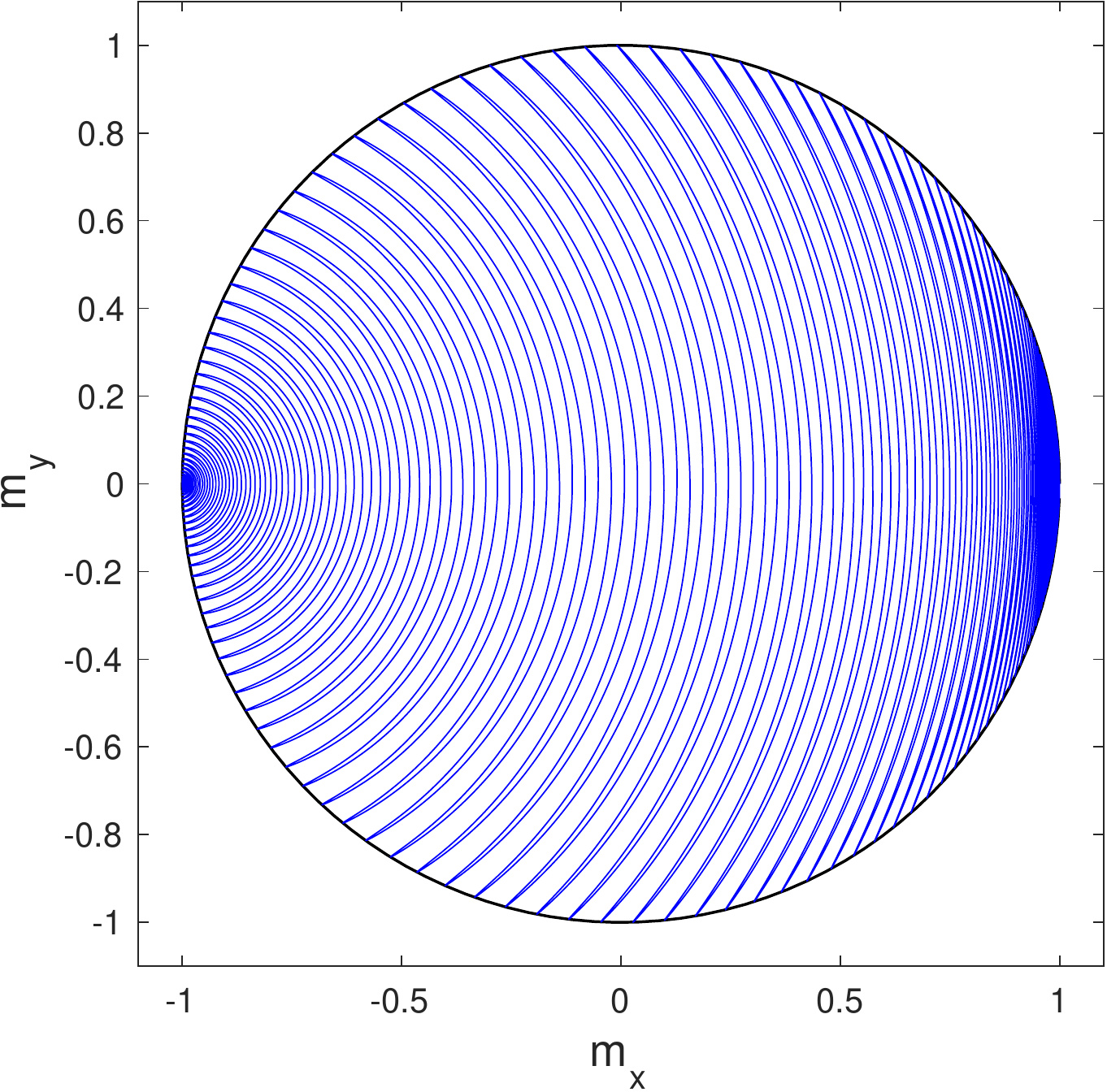}
        \caption{Small time step: 0.01}
        \label{fig:car_small}
    \end{subfigure}%
    ~ 
    \begin{subfigure}[t]{0.33\textwidth}
        \includegraphics[width=0.95\textwidth]{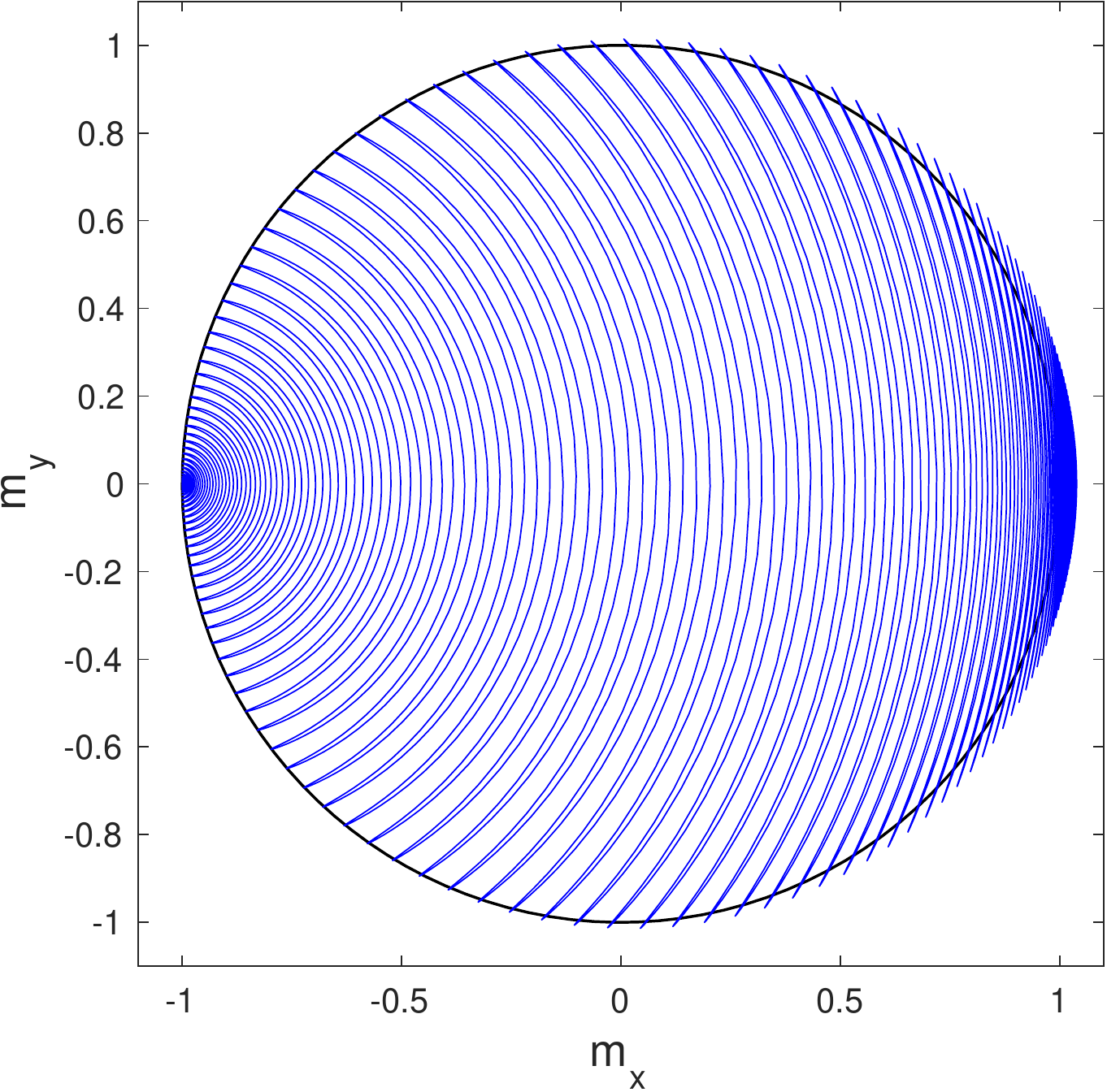}
        \caption{Moderate time step: 0.1}
        \label{fig:car_medium}
    \end{subfigure}%
    ~
    \begin{subfigure}[t]{0.33\textwidth}
        \includegraphics[width=0.95\textwidth]{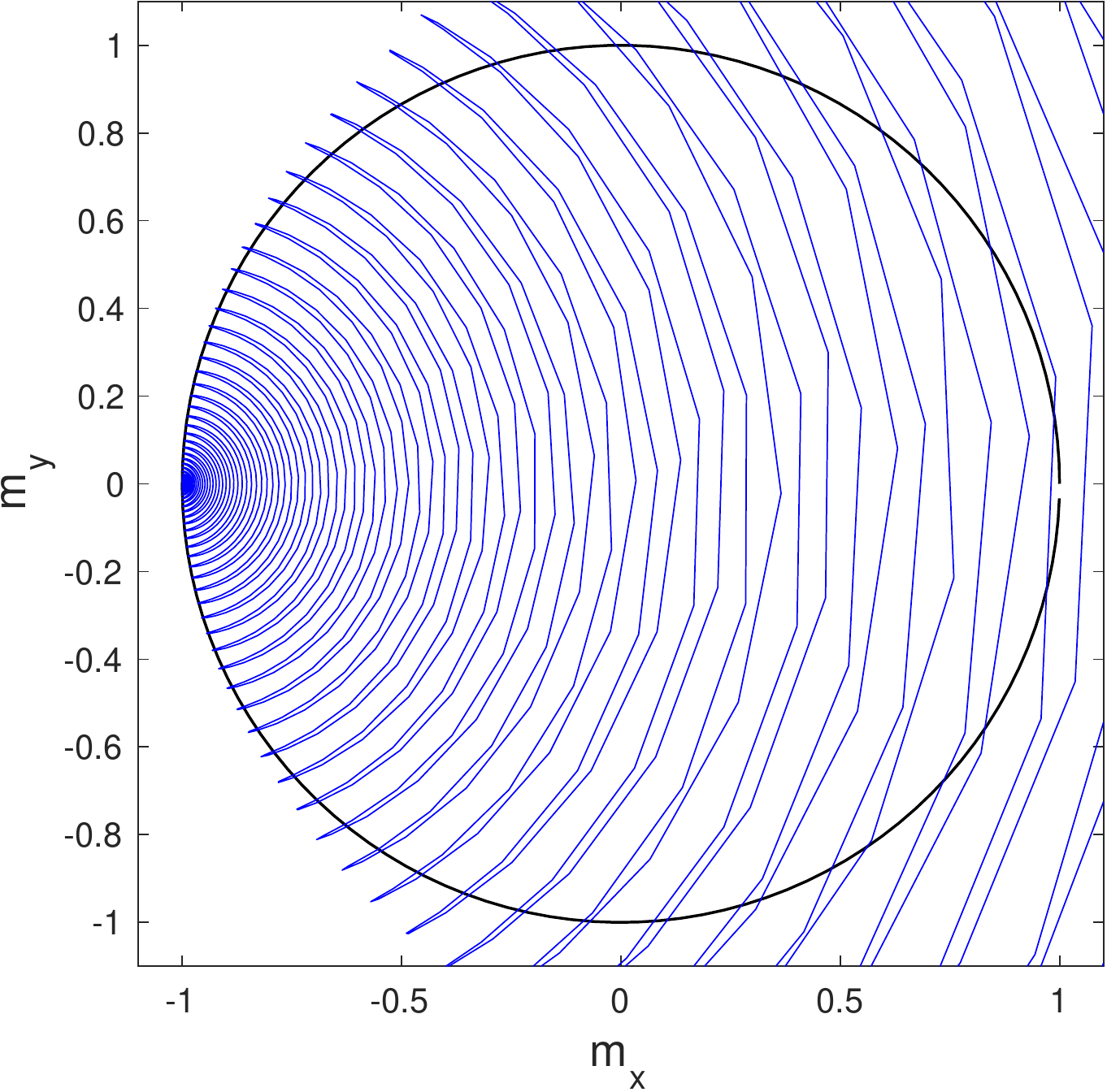}
        \caption{Large time step: 0.3}
        \label{fig:car_large}
    \end{subfigure}

    \caption{\footnotesize Cartesian form s-LLGS simulation of magnetization reversal from $-x$ to $+x$ direction for an applied field $H_{app}= M_s$ along positive x-axis for different time steps (normalized). The simulation parameters are same as that in Figure \ref{fig:spice_vs_matlab}. (a) Small time step yields both accurate and stable solution. (b) Moderate time step gives a stable solution which is stretched along the reversal direction, thereby underestimating the reversal delay (zero crossing of $m_x$). (c) Large time step results in unbounded and unstable solution.}
 \label{fig:carLLG}
\end{figure}

\begin{figure}[H]
    \begin{subfigure}[t]{0.33\textwidth}
 \includegraphics[width=0.95\textwidth]{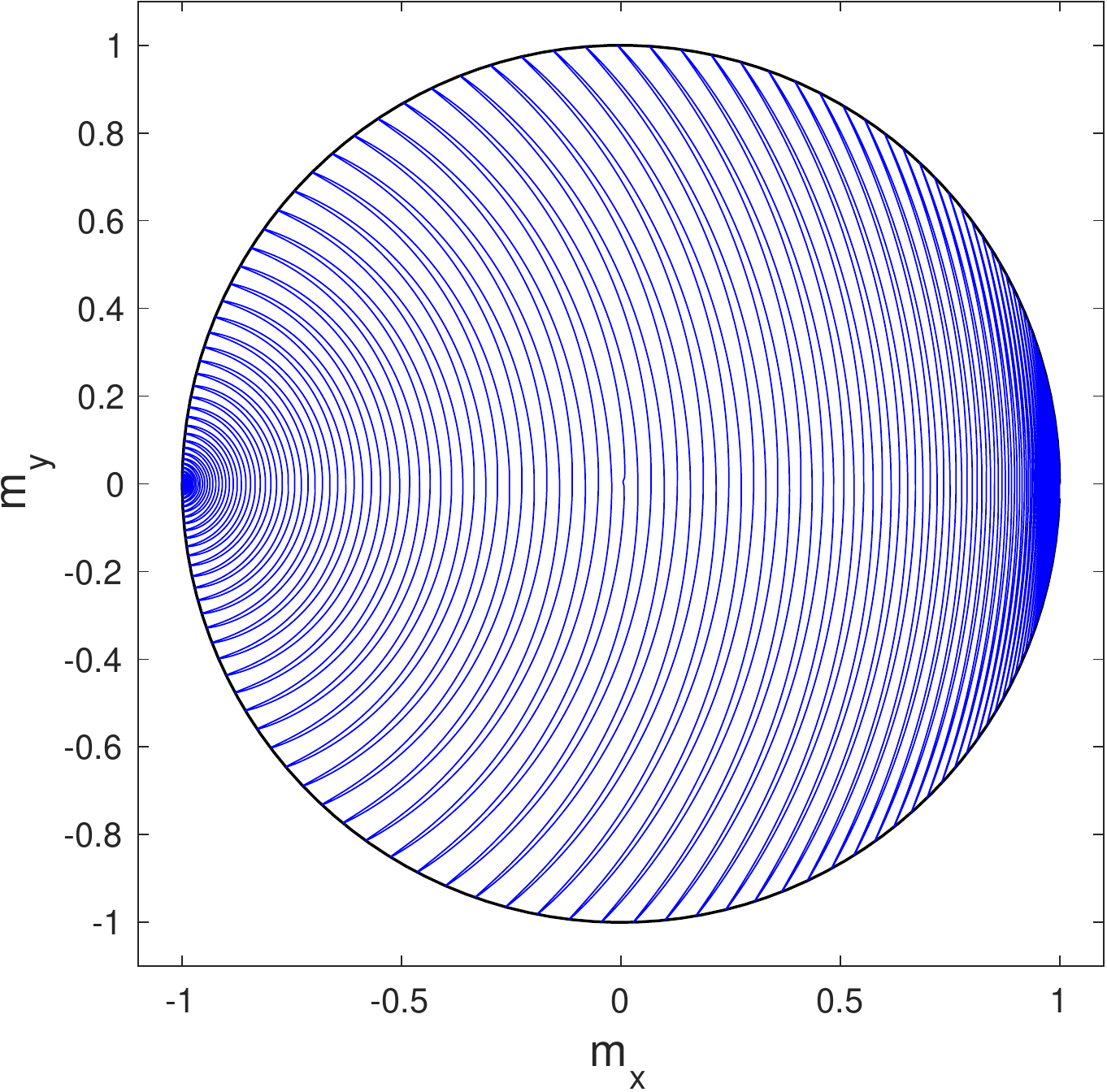}
        \caption{Small time step: 0.01}
        \label{fig:sph_small}
    \end{subfigure}%
    ~ 
    \begin{subfigure}[t]{0.33\textwidth}
        \includegraphics[width=0.95\textwidth]{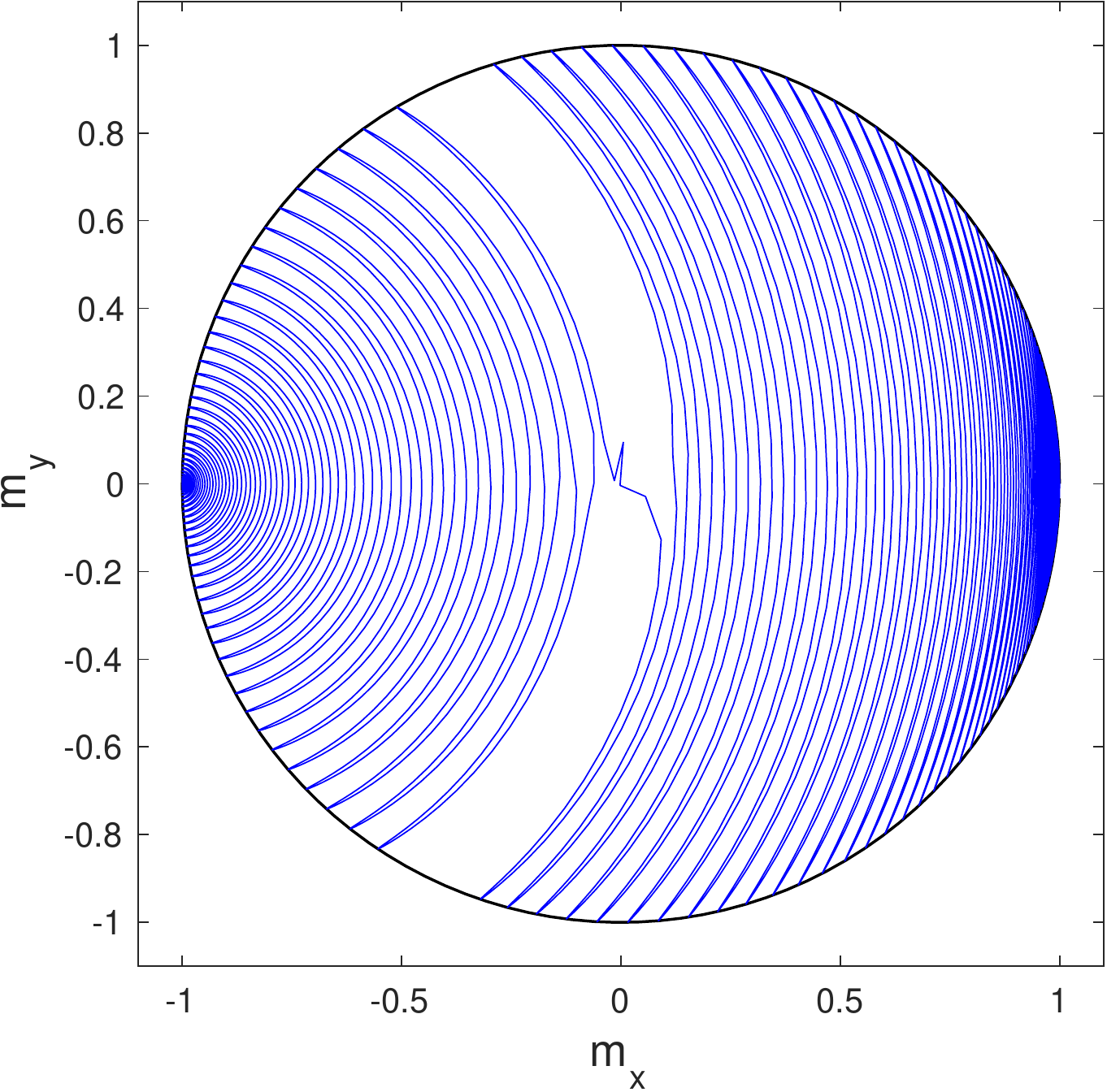}
        \caption{Moderate time step: 0.1}
        \label{fig:sph_medium}
    \end{subfigure}%
    ~
    \begin{subfigure}[t]{0.33\textwidth}
        \includegraphics[width=0.95\textwidth]{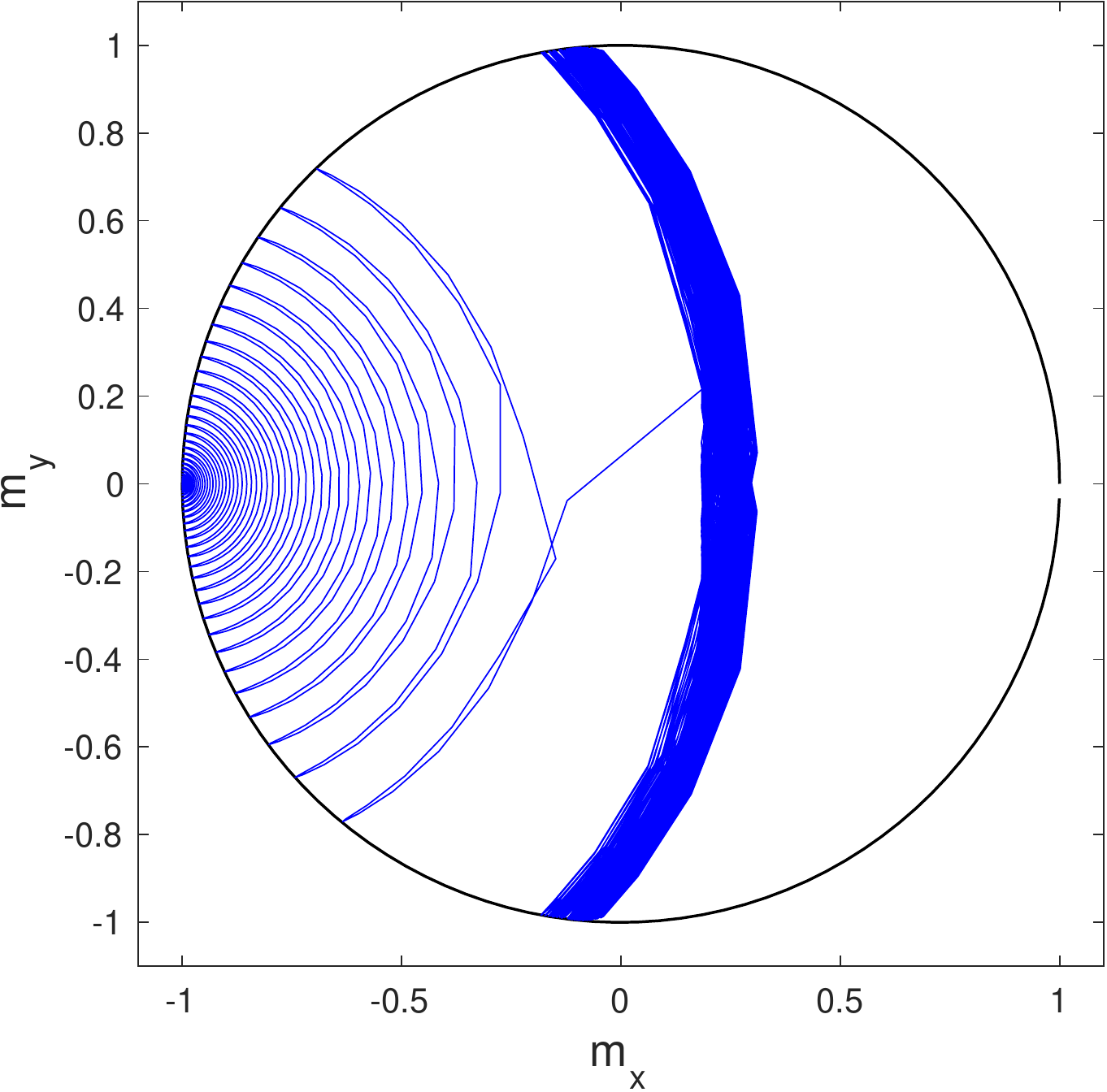}
        \caption{Large time step: 0.3}
        \label{fig:sph_large}
    \end{subfigure}

    \caption{\footnotesize Spherical form s-LLGS simulation of magnetization reversal from $-x$ to $+x$ direction for an applied field $H_{app}= M_s$ along positive x-axis for different time steps (normalized). The simulation parameters are same as that in Figure \ref{fig:spice_vs_matlab}. (a) Small time step yields both accurate and stable solution. (b) Moderate time step gives a stable solution with kinks and missing precessions around the unstable pole at $m_z=\pm 1$, and also underestimates the reversal delay. (c) Large time step results in bounded and oscillatory solution.}
 \label{fig:sphLLG}
\end{figure}

\noindent Figures \ref{fig:carLLG} and \ref{fig:sphLLG} show the trajectory of magnetization of the cartesian and spherical coordinates implementation of the s-LLGS equation for a range of time steps. For smaller time steps, the cartesian and spherical forms yield identical solutions which are both accurate and stable. For moderate time steps, the cartesian form stretches the trajectory to outside the unit circle; the spherical form has kinks around $m_z = \pm 1$ due to singularity of $\frac{d\phi}{dt}$ in (\ref{eq:LLG_sph}) at $\theta = 0, \pi$. This results in underestimation of reversal delay at moderate time step. For larger time steps, the cartesian and spherical forms produce unstable solutions, which are unbounded and oscillatory in nature respectively. In short, there is a trade-off between differentiability and boundedness in the cartesian and spherical implementations, respectively. 

\vspace{4ex}

\noindent We have already seen that time step is crucial to obtain an accurate and a stable solution. To get a smooth and continuous trajectory of $\mathbf{m}$, $|d\mathbf{m}|$ should be small for a given time step $dt$. Since, $\frac{d\mathbf{m}}{dt} \propto \mathbf{h}_{\mathrm{eff}}, \mathbf{i}_{\mathrm{s}}$, we get the constraint
\begin{equation}
\{|\mathbf{h}_{\mathrm{eff}}|, |\mathbf{i}_{\mathrm{s}}|\}dt\ll 1\quad \text{or} \quad 
\mathrm{max}\{|\mathbf{h}_{\mathrm{eff}}|, |\mathbf{i}_{\mathrm{s}}|\}dt\ll 1.
\label{eq:tscons1}
\end{equation}
From (\ref{heff_1}),  the magnitude of effective field is bounded by
\begin{align*}
|\mathbf{h}_{\mathrm{eff}}| &= \left|\mathbf{h}_{app} + \frac{H_k}{M_s}
(\hat{\mathbf{n}}\cdot \mathbf{m})\hat{\mathbf{n}} - \sum_i N_i m_i + \mathbf{h}_T\right| \\
&\leq |\mathbf{h}_{app}| + \left|\frac{H_k}{M_s}(\hat{\mathbf{n}}\cdot \mathbf{m})\hat{\mathbf{n}}\right| + \left|\sum_i N_i m_i \right| + |\mathbf{h}_T| \\ 
&=  |\mathbf{h}_{app}| + \frac{H_k}{M_s} + 1 + \nu = |\mathbf{h}_{app}| + \varepsilon
\end{align*}
where $\varepsilon = \frac{H_k}{M_s} + 1 + \nu$. Therefore, the constraint in (\ref{eq:tscons1}) becomes
\begin{equation}
\mathrm{max}\{|\mathbf{h}_{\mathrm{ap}}| + \varepsilon, |\mathbf{i}_{\mathrm{s}}|\}dt\ll 1.
\label{eq:tscons2}
\end{equation}
Suppose we take the LHS in (\ref{eq:tscons2}) to be 10 times smaller than 1, we get
\begin{equation}
dt_{\mathrm{crit}} = \frac{0.1}{\mathrm{max}\{|\mathbf{h}_{\mathrm{ap}}|+\varepsilon, |\mathbf{i}_{\mathrm{s}}|\}}.
\end{equation}
Note that the critical time step doesn't depend on $\alpha$, when we have chosen a $dt$ which satisfies the wider constraint in (\ref{eq:tscons1}). 
For the simulation results in Figures \ref{fig:car_small} and \ref{fig:sph_small}, we have $\varepsilon = 0.1+1+0 = 1.1$ and $|\mathbf{h}_{ap}| = 1$, which gives $\varepsilon = 2.1$, and $dt_{\mathrm{crit}} = 0.045$. Hence, we get a smooth trajectory in Figures \ref{fig:car_small} and \ref{fig:sph_small} since the time step chosen is smaller than this critical time step. But this is not the case for Figures 10(b-c) and 11(b-c), which results in unboundedness, discontinuities, and oscillations. 

\section{Benchmarking with OOMMF}
The Object Oriented MicroMagnetic Framework (OOMMF) by NIST is a widely used open source, portable, public domain tool for micromagnetics. The Oxs (OOMMF eXtensible Solver) is an extensible micromagnetic computation engine capable of solving problems defined on three-dimensional grids of rectangular cells holding three-dimensional spins \cite{Donahue}. We use OOMMF's Oxs with evolver class \textquotedblleft SpinXferEvolve" to simulate a single magnet whose magnetization evolves under the application of a spin current. Unfortunately there are no default evolver classes from OOMMF that include the effects of thermal noise. However, there are third-party extensions capableof performing thermal simulations in

\begin{figure}[H]
  \hspace*{-15ex}
    \includegraphics[scale=0.38]{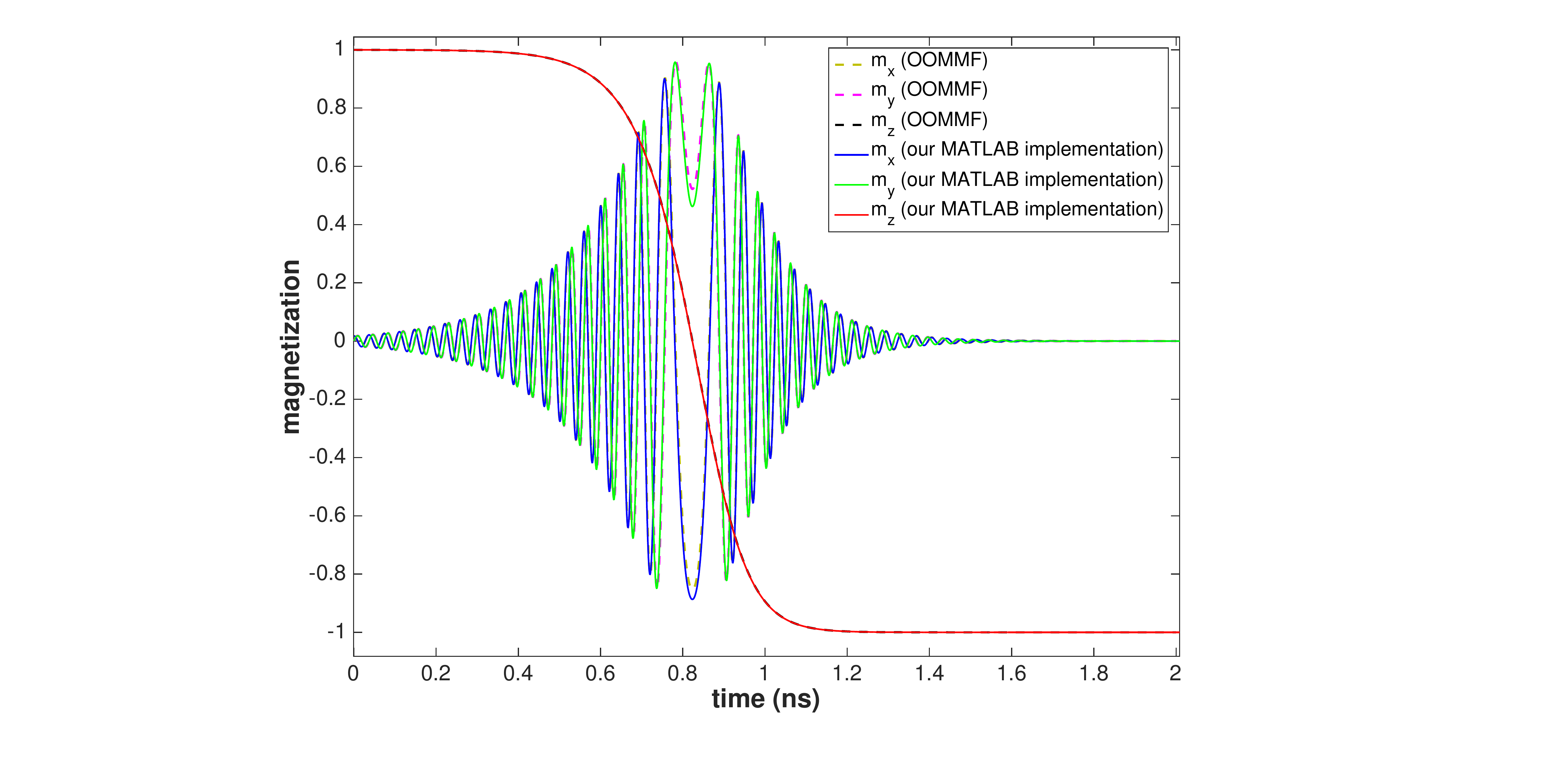}
  \vspace*{-5ex}
  \caption{\footnotesize The evolution of the magnetization vector under STT (without thermal effects) using our implicit midpoint implementation and the OOMMF solver. }
  \label{MATLABvsOOMMF}
\end{figure}

\noindent OOMMF. While the extension \textquotedblleft Xf$\_$ThermSpinXferEvolve" \cite{Fong} implements the solution of the s-LLGS equation using the Heun method, it naively renormalizes the magnetization norm in every iteration. The numerical accuracy and stability of this operation are questionable, and there appears to be no physical reason for this. In this section, we compare the magnetization vectors (from OOMMF) in the deterministic case with the corresponding vectors obtained from our implementation of the implicit midpoint method, as shown in Figure \ref{MATLABvsOOMMF}. This exercise substantiates the claim that our implicit midpoint realization is very close to the NIST standard.
 
\section{Outlook and Conclusion}
While a discussion of strong higher-order methods
was not the goal of this paper, there are notable developments,
whose applicability in solving the s-LLGS equation will be investigated in the future.
Namely, the efficient strong order stochastic Runge-Kutta schemes which
were proposed in \cite{Ro_2010_2}.
These schemes have the advantage that the number of evaluations
of the drift term of an SDE only increases linearly with the dimension
of the driving Wiener process. 
This is in contrast to previous methods, which relied on 
a quadratically increasing number of evaluations of the drift term.
For the approximation of the general multiple stochastic integrals, 
a method introduced in \cite{Wi_2001, Wi_2001_2} can be used.\\
\noindent Further, is important to note that many properties of magnets that are of interest 
to engineers can be solved for using weakly convergent schemes.
For example, the switching probability of a magnet at an arbitrary time 
is such a property.
Importantly, there are known schemes which converge in the weak sense with orders
up to 4 \cite{Kloeden_Platen}.
Recently, a very efficient explicit scheme with weak convergence order 2 
has been introduced in \cite{Ro_2009, Ro_2010}.
Many of these schemes can be
simplified so that no 
iterated stochastic integrals have to be computed.
This is a time-intensive process because
many random variables have to be computed
at each time step,
and it is one of the reasons why efficient 
high strong-order schemes are 
currently only available for equations
with special structure.
In order to take advantage of these higher order approximations, the authors are currently investigating 
weakly convergent higher-order schemes for the s-LLGS equation.\\

\noindent Summarising, in this paper, we examine the accuracy of (i)  implicit midpoint, (ii) Heun, (iii) Euler-Heun, and (iv) RK4-Heun numerical methods to solve the s-LLGS equation of macrospin dynamics. We compare the performance of these methods in terms of the average path-wise error, preservation of the magnetization norm, and 50\% switching boundary of macrospin reversal. We clearly show the higher accuracy of the implicit midpoint method as compared to other methods. However, the RK4-Heun method offers significant benefits in precision when the stochastic term in the s-LLGS equation is 2-3 orders of magnitude lower than the deterministic term. We also demonstrate that SPICE-based numerical integration schemes, such as trapezoidal, Gear, and Euler, when solving the cartesian s-LLGS equation, perform poorly and lead to significant errors in the results as compared to the implicit midpoint method at the same time step. We discuss the spherical coordinates implementation of the s-LLGS equation in SPICE as a possible solution to mitigate this problem, and also evaluate the critical time step required to obtain stable and accurate solutions for these implementations. Through an exhaustive study, we provide guidelines that will help researchers analyze macrospin dynamics more accurately in the context of appropriate stochastic calculus and the use of numerical integration method.
 
\section*{Acknowledgements}
\noindent The authors would like to acknowledge Prof. Supriyo Datta and Dr. Kerem Camsari from Purdue, and Nickvash Kani from Georgia Tech for the many useful discussions.

\appendix
\section{Derivation of dimensionless LLGS} \label{App:Appendix}
\noindent Beginning with \eqref{basic_sLLGS}, we divide both sides by $\gamma \mu_0M_s^2$,\\
\begin{equation}\frac{1}{\gamma \mu_0 M_s^2}\frac{d\bm{M}}{dt} = -\frac{1}{M_s^2} (\bm{M}\times \bm{H}_{eff}) + \frac{\alpha}{\gamma \mu_0M_s^3}\Big(\bm{M}\times \frac{d\bm{M}}{dt}\Big) - \frac{\bm{M}\times(\bm{M}\times \bm{I_s})}{q\gamma \mu_0M_s^3N_s}. \end{equation}\\

\noindent Let the scale for magnetization and effective field be $M_s$. Then normalizing as $\bm{m} = \frac{\bm{M}}{M_s}$ and $\bm{h}_{eff}  = \frac{\bm{H}_{eff}}{M_s}$,\\
\begin{equation}\frac{1}{\gamma \mu_0 M_s}\frac{d\bm{m}}{dt} = -\bm{m}\times \bm{h}_{eff} + \frac{\alpha}{\gamma \mu_0M_s}\Big(\bm{m}\times \frac{d\bm{m}}{dt}\Big) - \frac{\bm{m}\times(\bm{m}\times \bm{I_s})}{q\gamma \mu_0M_sN_s}. \end{equation}\\

\noindent Taking the current scale as $I = q\gamma \mu_0M_s N_s$, we obtain the normalized spin current $\bm{i_s} = \frac{\bm{I}_s}{I}$. Also we consider a new time scale $(\gamma \mu_0 M_s)^{-1}$ such that $t' = (\gamma\mu_0 M_s)t $ and $dt' = (\gamma\mu_0 M_s)dt.$\\
\begin{equation}\frac{d\bm{m}}{dt'} = -\bm{m}\times \bm{h}_{eff} + \alpha\Big(\bm{m}\times \frac{d\bm{m}}{dt'}\Big) - \bm{m}\times(\bm{m}\times \bm{i}_s).
\end{equation}\\
Changing variables $t := t'$ without loss of generality, we get
\begin{equation}
\label{eq_llg_implicit}
\frac{d\bm{m}}{dt} =  - \bm{m} \times \bm{h}_{eff} + \alpha \left( \bm{m} \times \frac{d\bm{m}}{dt} \right) - \bm{m} \times ( \bm{m} \times \bm{i}_s).\\
\end{equation}
This is the implicit form of the dimensionless s-LLGS equation.\\ 

\noindent To further transform this and decouple $\frac{d\bm{m}}{dt}$ from the right hand side of \ref{eq_llg_implicit}, we take the cross product with $\bm{m}$ on both sides,

\begin{equation}
\label{eq_llg_cross_m}
\bm{m}\times \frac{d\bm{m}}{dt} = -\bm{m}\times(\bm{m}\times \bm{h}_{eff}) + \alpha \bm{m}\times \Big(\bm{m}\times \frac{d\bm{m}}{dt}\Big) - \bm{m}\times (\bm{m}\times (\bm{m}\times \bm{i}_s)).  
\end{equation}

\noindent Now, 
\begin{equation}
\label{eq_llg_cross_m2}
\bm{m}\times\Big(\bm{m}\times \frac{d\bm{m}}{dt}\Big) = \bm{m}\Big(\bm{m}\cdot \frac{d\bm{m}}{dt}\Big) - \frac{d\bm{m}}{dt}(\bm{m}\cdot \bm{m}) = -|m|^2\frac{d\bm{m}}{dt},   
\end{equation}
 since $\bm{m}\cdot \frac{d\bm{m}}{dt} =0$.\\

\noindent And,
\begin{equation}
\label{eq_llg_cross_m3}
\begin{aligned}
\bm{m}\times (\bm{m}\times (\bm{m}\times \bm{i}_s)) &= \bm{m}(\bm{m} \cdot (\bm{m}\times \bm{i}_s)) - (\bm{m}\times \bm{i}_s)(\bm{m} \cdot  \bm{m}), \\
&= \bm{m}(\bm{i}_s \cdot (\bm{m}\times \bm{m})) - |m|^2(\bm{m}\times \bm{i}_s), \\
&= -|m|^2(\bm{m}\times \bm{i}_s).
\end{aligned}
\end{equation}

\noindent Substituting \eqref{eq_llg_cross_m2} and \eqref{eq_llg_cross_m3} back in equation \eqref{eq_llg_cross_m},
\begin{equation}
\label{eq_llg_cross_m4}
\bm{m}\times \frac{d\bm{m}}{dt} = -\bm{m}\times(\bm{m}\times \bm{h}_{eff}) - \alpha |m|^2\frac{d\bm{m}}{dt} + |m|^2(\bm{m}\times \bm{i}_s). 
\end{equation}\\
Replacing $\bm{m}\times \frac{d\bm{m}}{dt}$ from�~\eqref{eq_llg_cross_m4} in \eqref{eq_llg_cross_m},

\begin{equation}
\nonumber
\begin{aligned}
\frac{d\bm{m}}{dt} &= -\bm{m}\times \bm{h}_{eff} + \alpha\Big(-\bm{m}\times(\bm{m}\times \bm{h}_{eff}) - \alpha |m|^2\frac{d\bm{m}}{dt} + |m|^2(\bm{m}\times \bm{i}_s)\Big) - \bm{m}\times(\bm{m}\times \bm{i}_s) \\
 &= -\bm{m}\times \bm{h}_{eff} - \alpha \bm{m}\times(\bm{m}\times \bm{h}_{eff}) - \alpha^2 |m|^2\frac{d\bm{m}}{dt} + \alpha|m|^2(\bm{m}\times \bm{i}_s) - \bm{m}\times(\bm{m}\times \bm{i}_s), 
\end{aligned}
\end{equation}
which gives
\begin{equation}
(1+\alpha^2|m|^2)\frac{d\bm{m}}{dt} = -\bm{m}\times \bm{h}_{eff} - \alpha \bm{m}\times(\bm{m}\times \bm{h}_{eff}) + \alpha|m|^2(\bm{m}\times \bm{i}_s) - \bm{m}\times(\bm{m}\times \bm{i}_s).
\end{equation}
Now $|m|^2 = m_x^2 + m_y^2 + m_z^2 = 1$, so that
\begin{equation}
\label{eq_llg_explicit}
\frac{d\bm{m}}{dt} = - \frac{1}{1+\alpha^2} \left[ \bm{m} \times \bm{h}_{eff} +\bm{m} \times (\bm{m} \times \bm{i}_s) + \alpha \left( \bm{m} \times( \bm{m} \times \bm{h}_{eff}) -  \bm{m} \times \bm{i}_s \right)\right].
\end{equation}

\noindent This is the explicit form of the simensionless s-LLGS equation.\\ \\

\newpage

\bibliography{ArXiv_new}
\bibliographystyle{ieeetr}

\end{document}